\documentclass[bibyear]{aa}  

\usepackage{natbib}
\bibpunct{(}{)}{;}{a}{}{,} % to follow the A&A style

\usepackage{graphicx}

\newcommand{\rv}{R_\mathrm{V}}
\newcommand{\rmon}{R_\mathrm{0}}
\newcommand{\rc}{R_\mathrm{c}}

\newcommand{\xc}{x_\mathrm{c}}
\newcommand{\xmon}{x_\mathrm{0}}
\newcommand{\pmax}{P_\mathrm{max}}

\newcommand{\alphat}{\alpha_\mathrm{t}}

\newcommand{\org}{\texttt{org}}
\newcommand{\amc}{\texttt{amc}}

\usepackage{titlesec}
\titleformat{\subsubsection}[runin]% runin puts it in the same paragraph
       {\normalfont\bfseries}% formatting commands to apply to the whole heading
       {\thesubsection}% the label and number
       {0.5em}% space between label/number and subsection title
       {}% formatting commands applied just to subsection title
       [.]% punctuation or other commands following subsection title

\usepackage{txfonts}
\usepackage{color, xcolor}
\usepackage[colorlinks=true, linkcolor={blue!80!black},citecolor={blue!80!black},urlcolor={magenta}]{hyperref}

\begin{document} 

   \title{How large are the monomers of dust aggregates in planet-forming disks?}
   \subtitle{Insights from quantitative optical and near-infrared polarimetry}

   \author{R. Tazaki
          \inst{1,2}
          \and
          C. Dominik\inst{1}
          }

   \institute{Anton Pannekoek Institute for Astronomy, University of Amsterdam, Science Park 904, 1098XH Amsterdam, The Netherlands\\
              \email{r.tazaki@uva.nl}
         \and
             Astronomical Institute, Graduate School of Science, Tohoku University, 6-3 Aramaki, Aoba-ku, Sendai 980-8578, Japan\\
             }

   \date{}

% \abstract{}{}{}{}{} 
% 5 {} token are mandatory
 
  \abstract
  % context heading (optional)
   {
   The size of the constituent particles (monomers) of dust aggregates is one of the most uncertain parameters directly affecting collisional growth of aggregates in planet-forming disks. Despite its importance, the monomer size has not yet been meaningfully constrained by disk observations.
   }
  % aims heading (mandatory)
   {
   We attempt to derive the monomer size from optical and near-infrared (IR) polarimetric observations of planet-forming disks.
   }
  % methods heading (mandatory)
   {
   We perform a comprehensive parameter survey on the degree of linear polarization of light scattered by dust aggregates, using an exact numerical method called the $T$-matrix method. We investigate the effect of the monomer size, aggregate size, porosity, and composition on the degree of polarization. The obtained results are then compared with observed polarization fractions of several planet-forming disks at optical and near-IR wavelengths.
   }
  % results heading (mandatory)
   {
   It is shown that the degree of polarization of aggregates depends sensitively on the monomer size unless the monomer size parameter is smaller than one or two. Comparing the simulation results with the disk observations, we find that the monomer radius is no greater than $0.4~\mu$m. The inferred monomer size is therefore similar to subunit sizes of the solar system dust aggregates and the maximum size of interstellar grains.
   }
  % conclusions heading (optional), leave it empty if necessary 
   {
   Optical and near-IR quantitative polarimetry will provide observational grounds on the initial conditions for dust coagulation and thereby planetesimal formation in planet-forming disks.
   }

   \keywords{Scattering --
                Polarization --
                Protoplanetary disks
               }

    \maketitle
%-------------------------------------------------------------------
\section{Introduction}
Collisional growth of dust aggregates is the first step of planet formation. 
The properties of the constituent particles (monomers) of an aggregate are important as they directly affect the impact strength of the aggregate, which intimately links to the maximum aggregate radius to be reached by collisional growth \citep{Birnstiel12, Okuzumi16,  Pinilla17, Okuzumi19}. In particular, the monomer size and surface materials coating each monomer are of crucial relevance for the impact strength. For example, larger monomers and/or less sticky materials make aggregates more fragile \citep{Chokshi93, Dominik97, Wada09}. Although material dependence of the impact strength has been extensively studied by laboratory experiments in the last decades \citep[e.g.,][]{Poppe00, Gundlach15, Musiolik16a, Musiolik16b}, the monomer size remains uncertain.

Previous studies of dust coagulation commonly assume sub-micron-sized monomers \citep{Weidenschilling84, Weidenschilling97, Dullemond05, Ormel07, Okuzumi12, Kataoka13, Krijt16, Kobayashi21}. This monomer size has been motivated by previous studies on the subunit sizes of interplanetary dust particles \citep{Brownlee85, Rietmeijer93, Woz13} or the largest size of interstellar grains \citep{MRN77, Draine84, Jones13}. Recent in situ measurements of dust aggregates of the comet 67P/Churyumov--Gerasimenko (C-G) also suggest the presence of sub-micron-sized subunits \citep{Bentley16, Mannel16, Mannel19}. In this way, the current estimates of the monomer radius have relied on perspectives on the solar system or the interstellar medium.

Despite its essential importance in collisional growth, the monomer radius has not yet been meaningfully constrained by observations of planet-forming disks. \citet{Graham07} found that optical scattered light from the debris disk around AU Mic is highly polarized, hinting at the presence of porous aggregates of sub-micron monomers; however, the authors mainly focused on the porosity of the aggregates rather than the size of the monomers. 
In contrast, recent observations of younger disks start to question the presence of sub-micron-sized monomers. The detections of millimeter-wave scattering polarization seem to imply that collisional growth stalls at a small aggregate size ($\sim100~\mu\mathrm{m}$) \citep{Kataoka16, Stephens17}, which may be attributed to fragile nature of aggregates, such as due to micron-sized (or even larger) monomers and/or less sticky materials \citep{Okuzumi19, Arakawa21}. 

Observations of polarized scattered light from planet-forming disks would be crucial to clarify the monomer size.
The scattering polarization properties of large porous aggregates have been suggested to strongly reflect the properties of the monomer, as reported in numerical simulations \citep{West91, Kozasa93, Lumme97, Petrova00, Petrova04, Kimura01, Kimura06, Bertini07, Kolokolova10, Tazaki16, Min16, Halder18} and laboratory experiments \citep{Zerull93, Gustafson99, Volten07}. In contrast, \citet{Shen09} pointed out that the effect of monomer size on polarization is negligible for aggregates with a porosity of $\sim60\%$. However, they only investigated small monomer sizes, and hence, it is unclear whether their conclusions remain valid for larger monomers. In the solar system, optical polarimetry of comets has successfully revealed the presence of sub-micron-sized monomers in cometary aggregates \citep{Gustafson99, Kimura03, Kimura06}. 

This study aims to investigate the monomer size from a new vantage point, that is, polarimetric observations of planet-forming disks from optical to near-infrared (IR) wavelengths. To our knowledge, this is the first attempt at deriving the monomer radius from disk observations. As shown later in this paper, optical and near-IR wavelengths are the optimal wavelengths for distinguishing sub-micron- and micron-sized monomers. 

\begin{figure}[t]
\begin{center}
\includegraphics[width=0.96\linewidth]{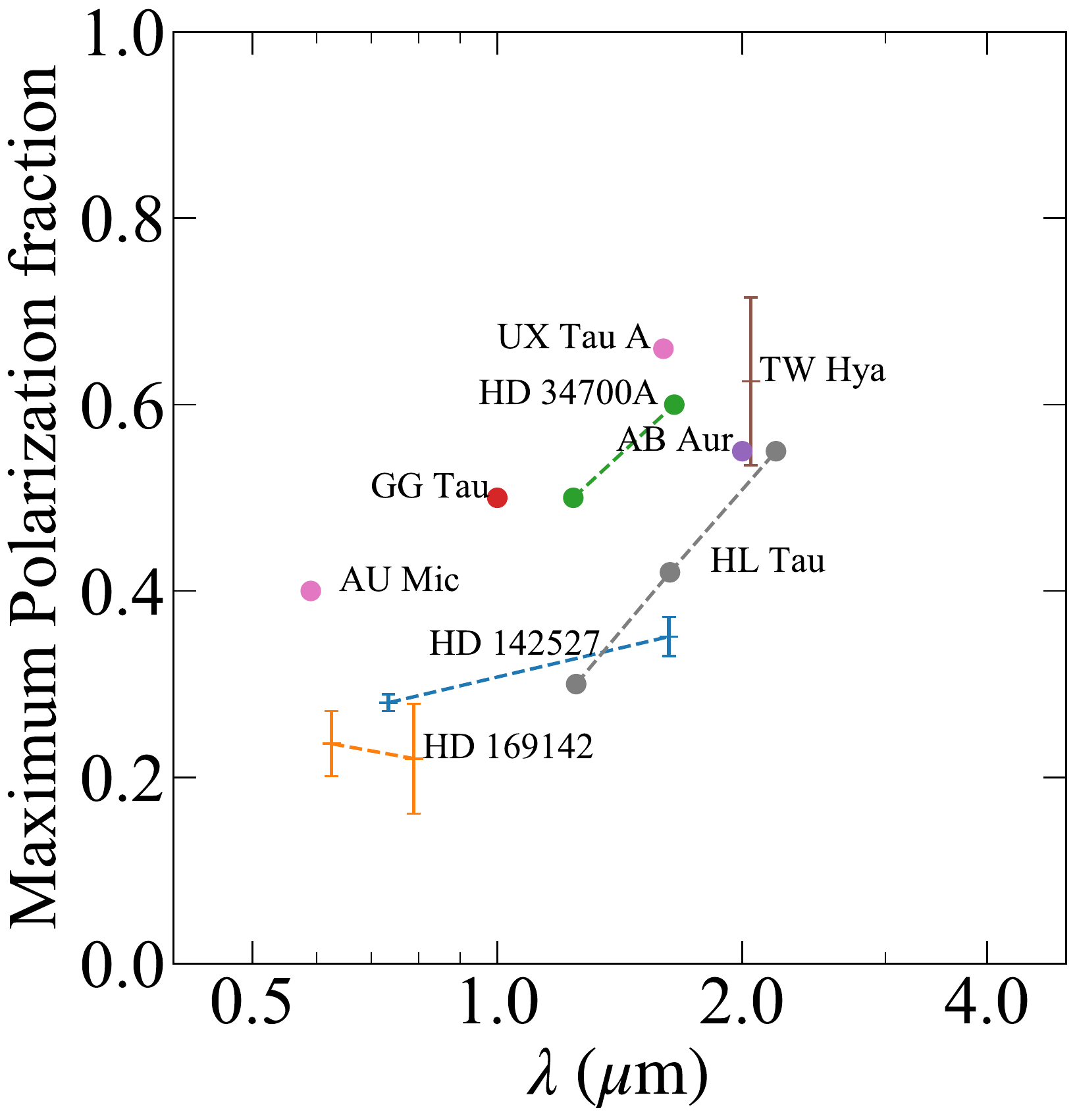}
\caption{Observed maximum polarization fractions $P_\mathrm{max}^\mathrm{obs}$ of several planet-forming disks. References: HD 142527 \citep{Hunziker21}, HD 169142 \citep{Tschudi21}, HD 34700 A \citep{Monnier19}, GG Tau \citep{Silber00}, AB Aur \citep{Perrin09}, UX Tau A \citep{Tanii12}, TW Hya \citep{Poteet18}, HL Tau (nebula region) \citep{Murakawa08} and AU Mic \citep{Graham07}.}
\label{fig:pobs}
\end{center}
\end{figure}

Fig. \ref{fig:pobs} compiles the maximum values of the observed polarization fractions ($P_\mathrm{max}^\mathrm{obs}$) of several planet-forming disks, where the polarization fraction means the degree of linear polarization of the observed polarization signals from the disk.
Except for an edge-on debris disk (AU Mic), we generally have access to light reflected off the disk with a scattering angle of 90 degrees at which the degree of linear polarization of aggregates is often maximized \citep[e.g.,][]{Kimura06, Volten07, Min16}. Thus, the observed maximum polarization fraction should be correlated with the maximum degree of polarization of light scattered by each aggregate ($\pmax$) in those disks. We can also say that the maximum degree of polarization of aggregates must be higher than the observed maximum polarization fractions ($\pmax\gtrsim P_\mathrm{max}^\mathrm{obs}$) because disk scattered light usually suffers a depolarizing effect due to multiple scattering (for an optically thick disk), line-of-sight integration (for an optically thin edge-on disk, such as AU Mic), or limited spatial resolution. Considering the presence of such depolarization effects, we can rule out the aggregate models that yield $\pmax<P_\mathrm{max}^\mathrm{obs}$. 

As the number of disks with quantitative polarimetry data is currently limited, we shall adopt a working hypothesis that the monomer size and composition are the same among those disks, but the aggregate size and porosity may vary from one disk to another. With the hypothesis, aggregates with a successful monomer model should be capable of producing the degree of polarization of $\gtrsim24-40\%$ and $\gtrsim35-66\%$ at optical and near-IR wavelengths, respectively, as well as a  reddish polarization color, i.e., $P_\mathrm{max}^\mathrm{obs}$(optical)$\lesssim P_\mathrm{max}^\mathrm{obs}$({near-IR}).

The main goal of this paper is to clarify what kind of aggregates and monomers can meet these criteria. To this end, we calculate $\pmax$ values of various kinds of aggregates by performing exact light scattering simulations. Comparing the obtained $\pmax$ values with $P_\mathrm{max}^\mathrm{obs}$, we demonstrate that aggregates are likely consisting of monomers less than or comparable to 0.4$~\mu$m in radius, while a larger monomer model fails to explain the observations. 

The paper is organized as follows. Sect. \ref{sec:model} summarizes our aggregate models and a method for solving light scattering. The outcomes of the light scattering simulations are presented in Sect. \ref{sec:result}. In Sect. \ref{sec:142527}, we apply the simulation results to recent quantitative polarimetric observations of the disk around HD 142527. In Sect. \ref{sec:discussion}, we discuss the results and finally in Sect. \ref{sec:summary} we summarize our findings.

%--------------------------------------------------------------------
\section{Models and Methods} \label{sec:model}

Dust coagulation in a planet-forming disk starts with hit-and-stick collisions, which have often been modeled by ballistic cluster cluster aggregation (BCCA) or ballistic particle cluster aggregation (BPCA). These aggregation models tend to produce fluffy aggregates whose porosity is higher than $\sim85\%$. As they grow, the aggregates may experience bouncing collision \citep{Zsom10, Lorek18}, although it is still a matter of debate if fluffy aggregates are eligible for bouncing collision \citep{Wada11, Seizinger13}. If bounce collisions occur, the aggregates suffer gradual compaction after each collision, and then the porosity will decrease to $\sim64~\%$ \citep{Weidling09}. 

\begin{figure}[t]
\begin{center}
\includegraphics[width=0.96\linewidth]{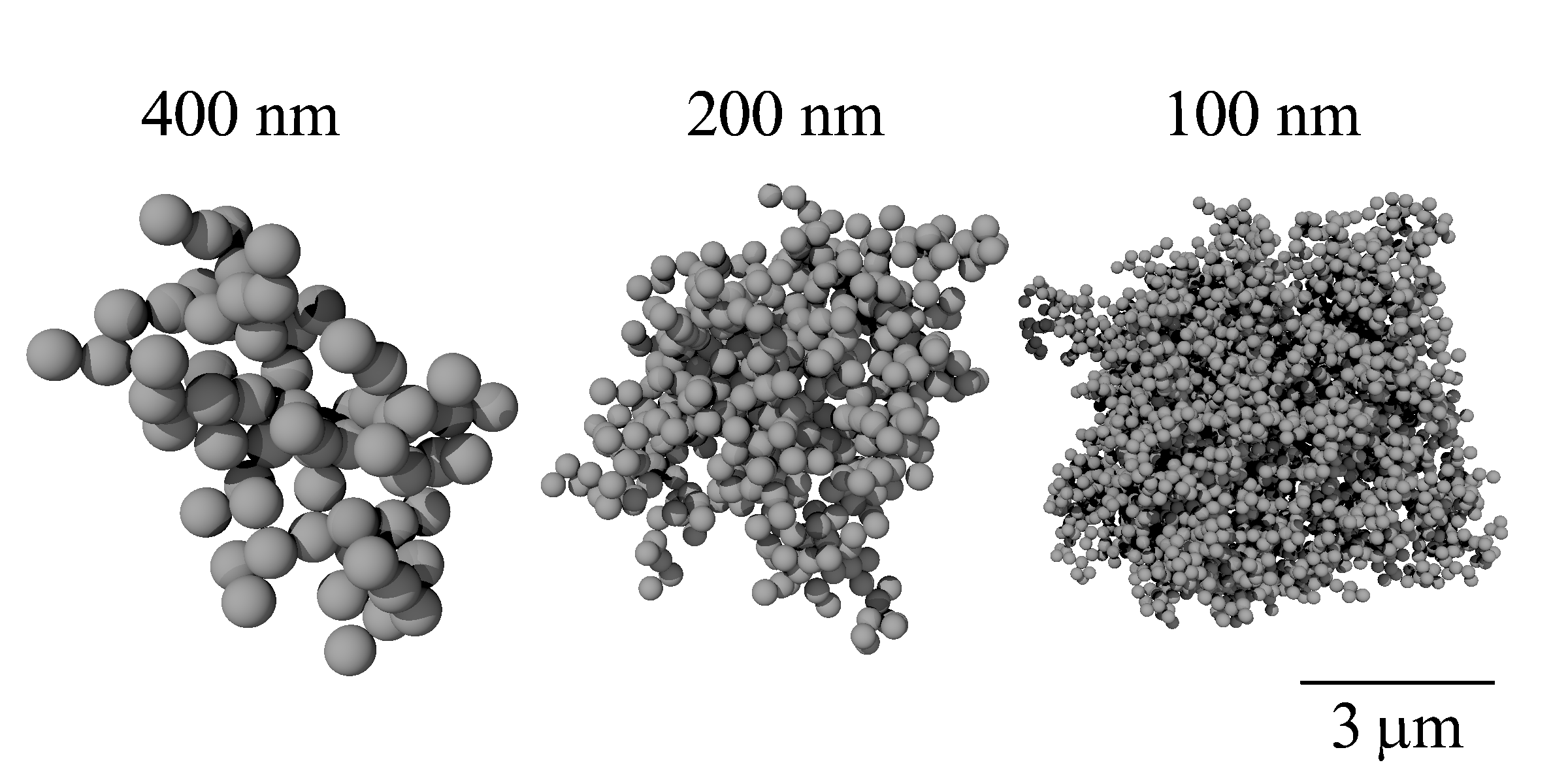}
\includegraphics[width=0.96\linewidth]{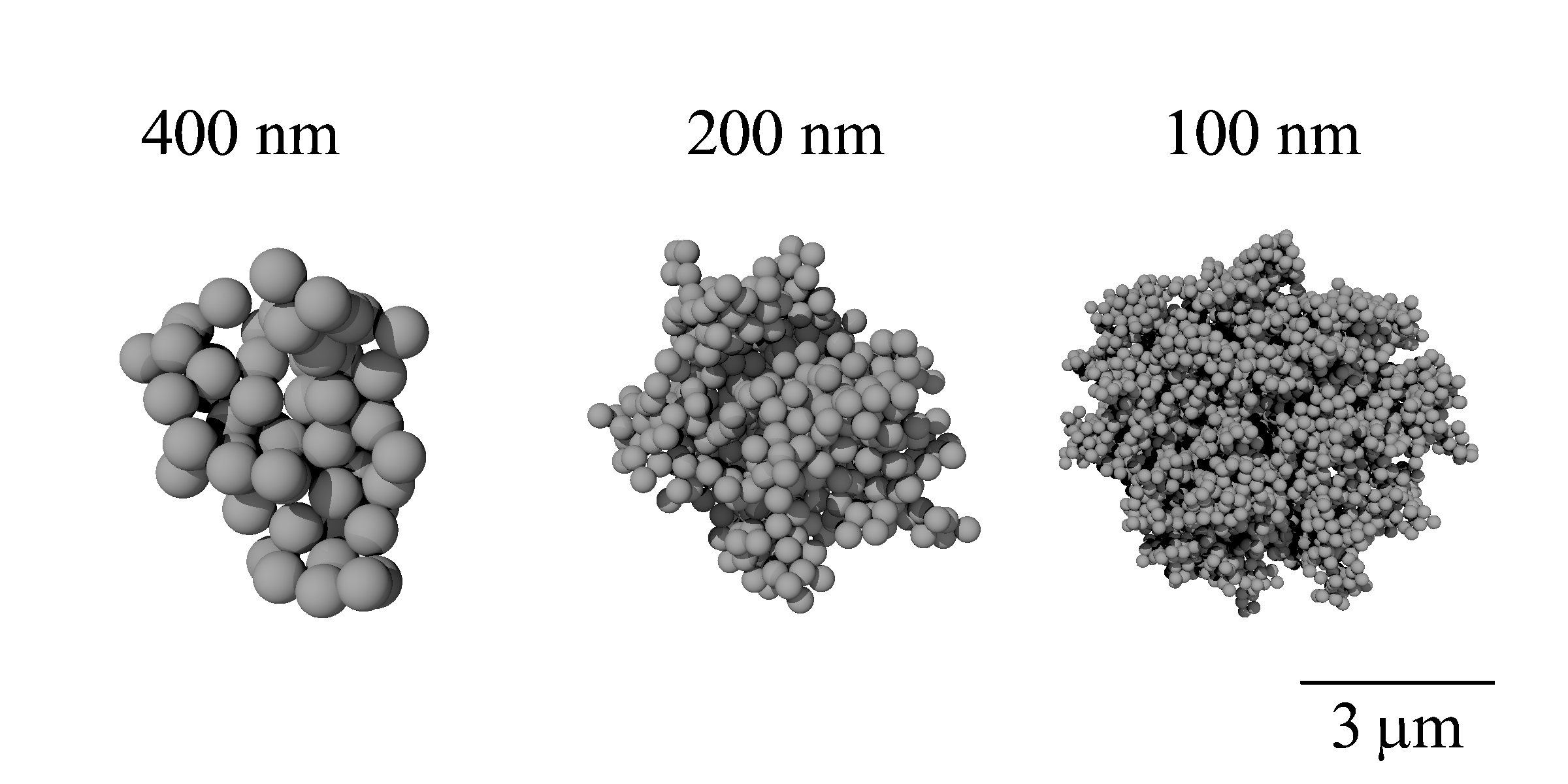}
\includegraphics[width=0.96\linewidth]{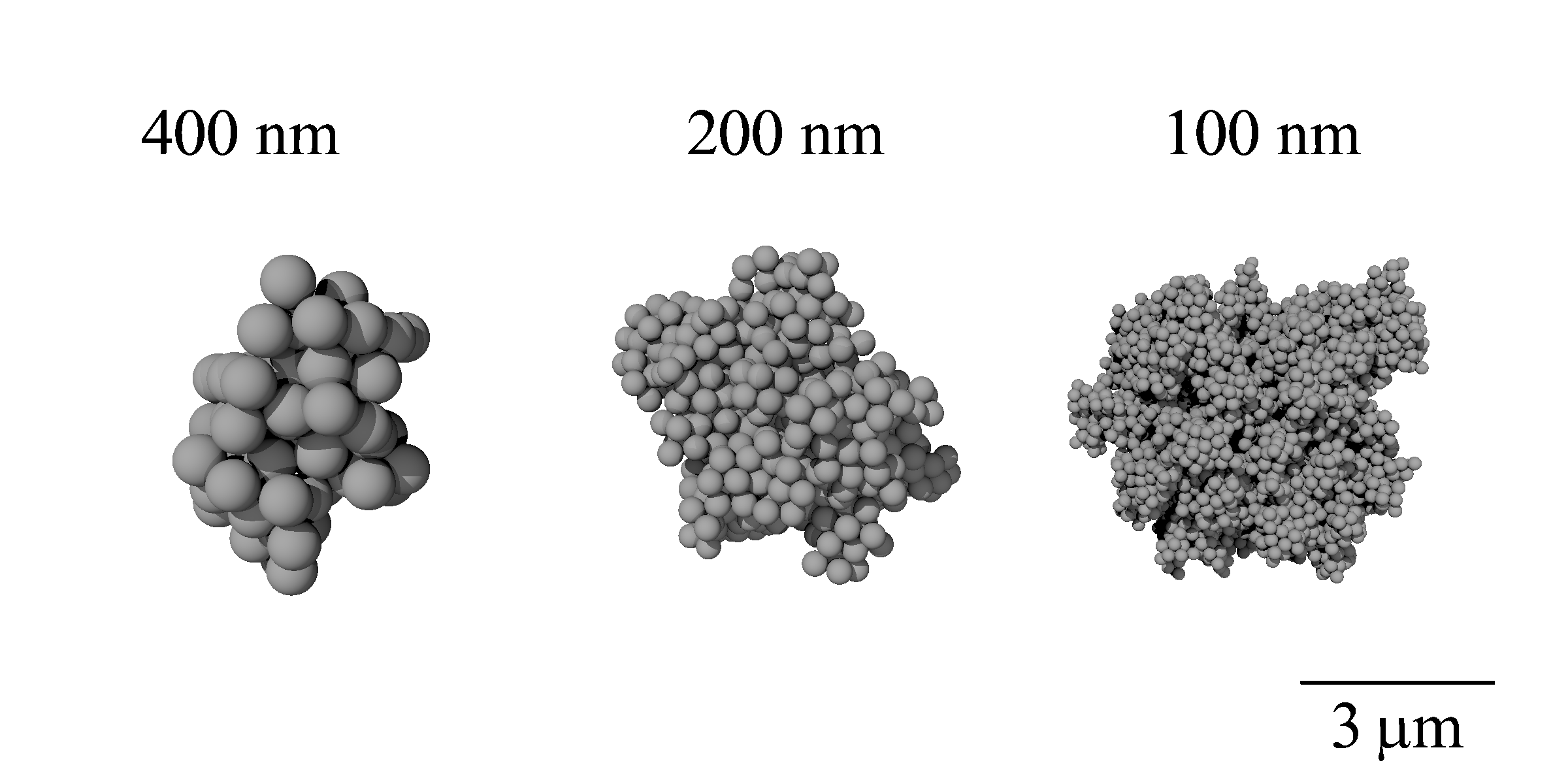}
\caption{BPCA ({\it upper}), BAM1 ({\it middle}), and BAM2 ({\it lower}) clusters with the volume-equivalent radius of $\rv=1.6~\mu$m. From left to right, the number of monomers and the monomer radius ($N$, $\rmon$) are ($64$, 400 nm), ($512$, 200 nm), and ($4096$, 100 nm), respectively.}
\label{fig:agg}
\end{center}
\end{figure}

With these in mind, we adopt three types of aggregates built from mono-disperse spherical monomers \footnote{The particle position data is available from B. T. Draine's Web site \url{https://www.astro.princeton.edu/~draine/agglom.html}.}. The first aggregation model is BPCA, where a single monomer particle is shot one-by-one to the target aggregate. BPCA clusters typically have a fractal dimension of $\sim3$ and a porosity of $85$--$87\%$ at $N\ge32$, where $N$ is the number of monomers.
Since the degrees of polarization of BCCA and BPCA clusters are more or less similar \citep{Kolokolova06, Kimura06}, we focus only on BPCA as a representative of hit-and-stick agglomerates. 
To investigate aggregates with lower porosity, we also consider aggregation models called BAM1 and BAM2 \citep{Shen08}. BAM1 and BAM2 are a modified version of BPCA. The first contact point between an arriving monomer and the target aggregate was found similarly to BPCA, but BAM1 and BAM2 allow the newly attached monomer to roll over the aggregate surface to find the second and third contact points, respectively. In this way, the resultant aggregates become less porous than BPCA clusters, i.e., a porosity of $\sim 69$--$78\%$ for BAM1 and $\sim47$--$68\%$ for BAM2 at $N\ge32$. The BAM2 clusters have a similar porosity to compressed aggregates with multiple bouncing collisions \citep{Weidling09}. The aggregate shapes are shown in Fig. \ref{fig:agg}.

To measure the size of an aggregate, we use the volume-equivalent radius $\rv=\rmon N^{1/3}$, where $\rmon$ is the monomer radius. 
For example, if two aggregates have the same $\rv$ value, they have the same material volume. As the volume-equivalent radius usually does not represent an apparent size of an aggregate, we also calculate the characteristic radius $\rc=\sqrt{5/3}R_g$ as a measure of its apparent size, where $R_g$ is the radius of gyration \citep{Kozasa92, Mukai92}. 
Namely, $\rv$ and $\rc$ specify the mass and apparent size of an aggregate, respectively.
The size parameters of the monomer and the aggregate are defined by $\xmon=2\pi \rmon/\lambda$ and $\xc=2\pi \rc/\lambda$, respectively, where $\lambda$ is a wavelength. Also, to measure the porosity of an aggregate, we adopt a commonly used definition: $\mathcal{P}=1-(\rv/\rc)^3$. 

This study adopts three different monomer radii: $\rmon=0.1~\mu$m, $0.2~\mu$m, and $0.4~\mu$m. 
The number of monomers is set as $N=2^i$, where $i$ runs from $3$ to $i_\mathrm{max}$ with an increment of one, and $i_\mathrm{max}=12$, $9$, $6$ for $\rmon=0.1~\mu$m, $0.2~\mu$m, and $0.4~\mu$m, respectively.
Thus, the largest aggregates we studied have $\rv=1.6~\mu$m, corresponding to $\rc=2.3$, $2.6$, and $3.1~\mu$m for BAM2, BAM1, and BPCA clusters with $\rmon=0.1~\mu$m, respectively. 
These sizes are large enough to study aggregates at the disk surfaces probed by optical and near-IR observations, as larger aggregates tend to settle down below the disk scattering surfaces.

As a monomer composition, we consider a mixture of pyroxene silicate (Mg$_{0.7}$Fe$_{0.3}$SiO$_3$) \citep{Dorschner95}, water ice \citep{Warren08}, carbonaceous material (either organics or amorphous carbon), and troilite \citep{Henning96} with the mass fractions provided by \citet{Birnstiel18}. The mass abundance of water ice adopted in \citet{Birnstiel18} was reduced from its base models \citep{Pollack94, dalessio01}. Such a reduced water-ice abundance is in line with recent modelings of water-ice features arising from the surface regions of the disks \citep{Tazaki21, Betti22}. 
Since the actual form of carbonaceous material is rather uncertain, we also consider two carbonaceous materials: one is organics \citep{Henning96}, and the other one is amorphous carbon \citep{Zubko96}. 

We derive an effective refractive index for a mixture of the four materials using the Bruggeman mixing rule. The resultant optical constants are summarized in Table \ref{tab:opcont}. The monomers comprising organics have refractive indices lower than those comprising amorphous carbon. We will refer to each model as the \org~ model and \amc~ model. To specify the monomer size and composition, we will use a model name in a form of \texttt{COMP}-\texttt{SIZE}, where \texttt{COMP} and \texttt{SIZE} specify the monomer composition (\org~or \amc) and radius in unit of nm (\texttt{100}, \texttt{200}, or \texttt{400}), respectively. For example, \amc-\texttt{200} indicates a monomer with the $\amc$ composition and a radius of 200 nm. We also tested the effect of ice-mantled monomers on the polarization characteristics of aggregates and found no significant differences between ice-mantled and fully mixed cases. Thus, in the following, we only consider homogeneous monomers for simplicity.

\begin{table}[t]
\caption{Real ($n$) and imaginary ($k$) parts of the refractive index of each monomer made of a mixture of silicate, water ice, carbonaceous (organics or amorphous carbon), and troilite.}
\label{tab:opcont}
\centering
\begin{tabular}{cccccc} 
\hline\hline          
model & \multicolumn{2}{c}{\org}  && \multicolumn{2}{c}{\amc}\\ 
\cline{2-3}
\cline{5-6}
$\lambda$ ($\mu$m) & $n$ & $k$ & & $n$ & $k$ \\
\hline
 3.78 & 1.53 & 0.0219 && 2.13 & 0.393 \\
 2.18 & 1.47 & 0.0134 && 1.98 & 0.385 \\
 1.63 & 1.48 & 0.0138 && 1.92 & 0.404 \\
 1.25 & 1.49 & 0.0104 && 1.86 & 0.420 \\
 1.04 & 1.49 & 0.0108 && 1.81 & 0.434 \\
0.735 & 1.50 & 0.0119 && 1.70 & 0.468 \\
0.554 & 1.51 & 0.0138 && 1.59 & 0.472 \\
\hline
\end{tabular}
\end{table}   

To solve light scattering by dust aggregates, we use an exact numerical technique so-called the cluster $T$-matrix method \citep[][and references therein]{Mackowski96}. We use a publicly available code for the cluster $T$-matrix method \texttt{MSTM~v3.0} \citep{Mackowski11}. The great advantage of the $T$-matrix approach is its analyticity of mathematical formulation, which allows analytical orientation averaging. All cluster $T$-matrix simulations presented in this paper adopt analytical orientation averaging. 
The results are also averaged over 4 realizations for each aggregate model. Because the degree of linear polarization does not vary significantly between realizations, increasing the number of realizations furthermore would not change our conclusions.

\section{Results} \label{sec:result}

We present the results of cluster $T$-matrix simulations. First of all, we study the effect of monomer size on the degree of linear polarization of light scattered by aggregates in Sect. \ref{sec:mon}. In Sect. \ref{sec:param}, we summarize the results of our comprehensive parameter survey on the polarization properties, which will then be compared with the observed polarization fractions in Sect. \ref{sec:obs} to derive the monomer radius in planet-forming disks.

\begin{figure}[t]
\begin{center}
\includegraphics[width=0.96\linewidth]{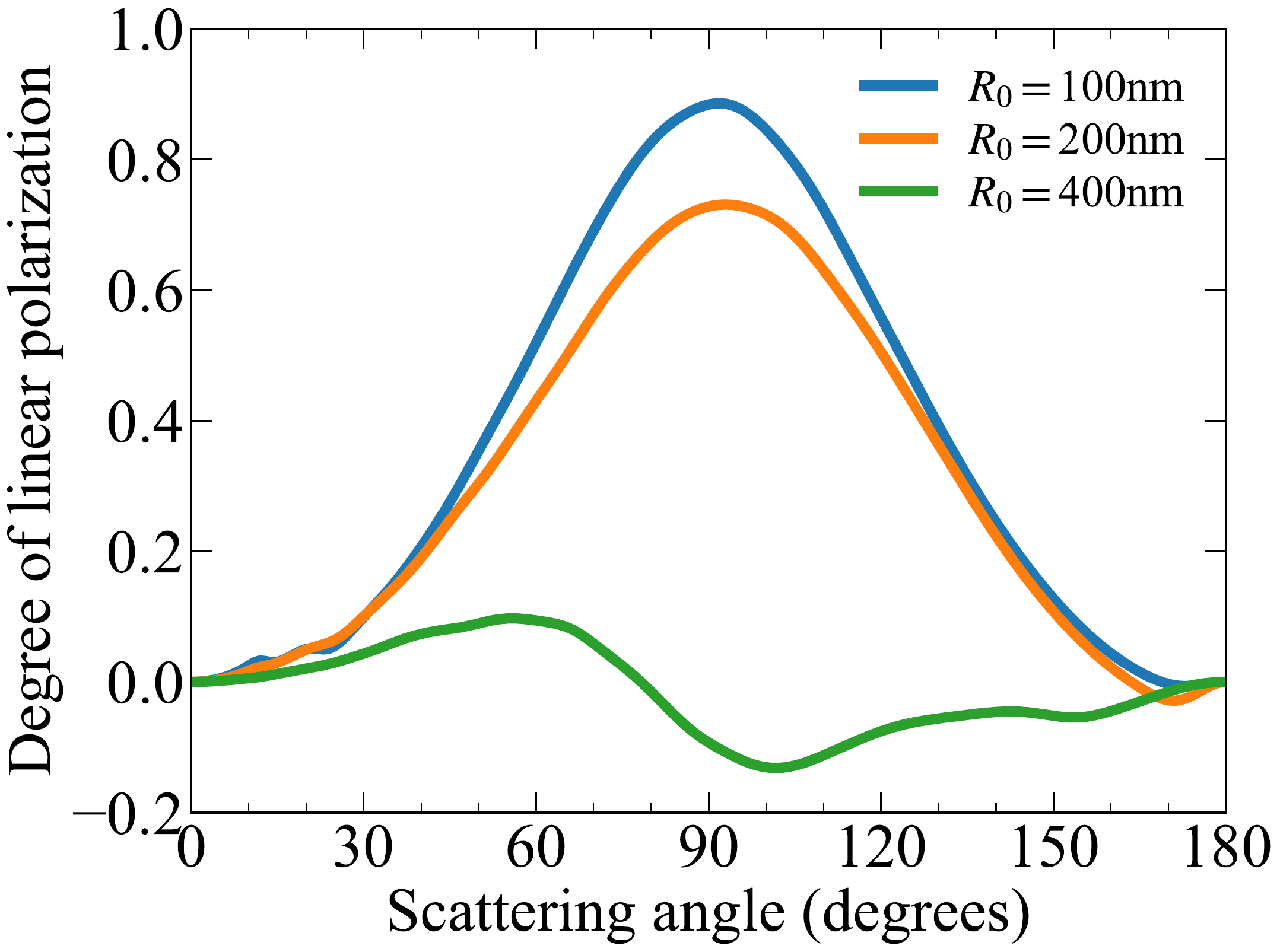}
\includegraphics[width=0.96\linewidth]{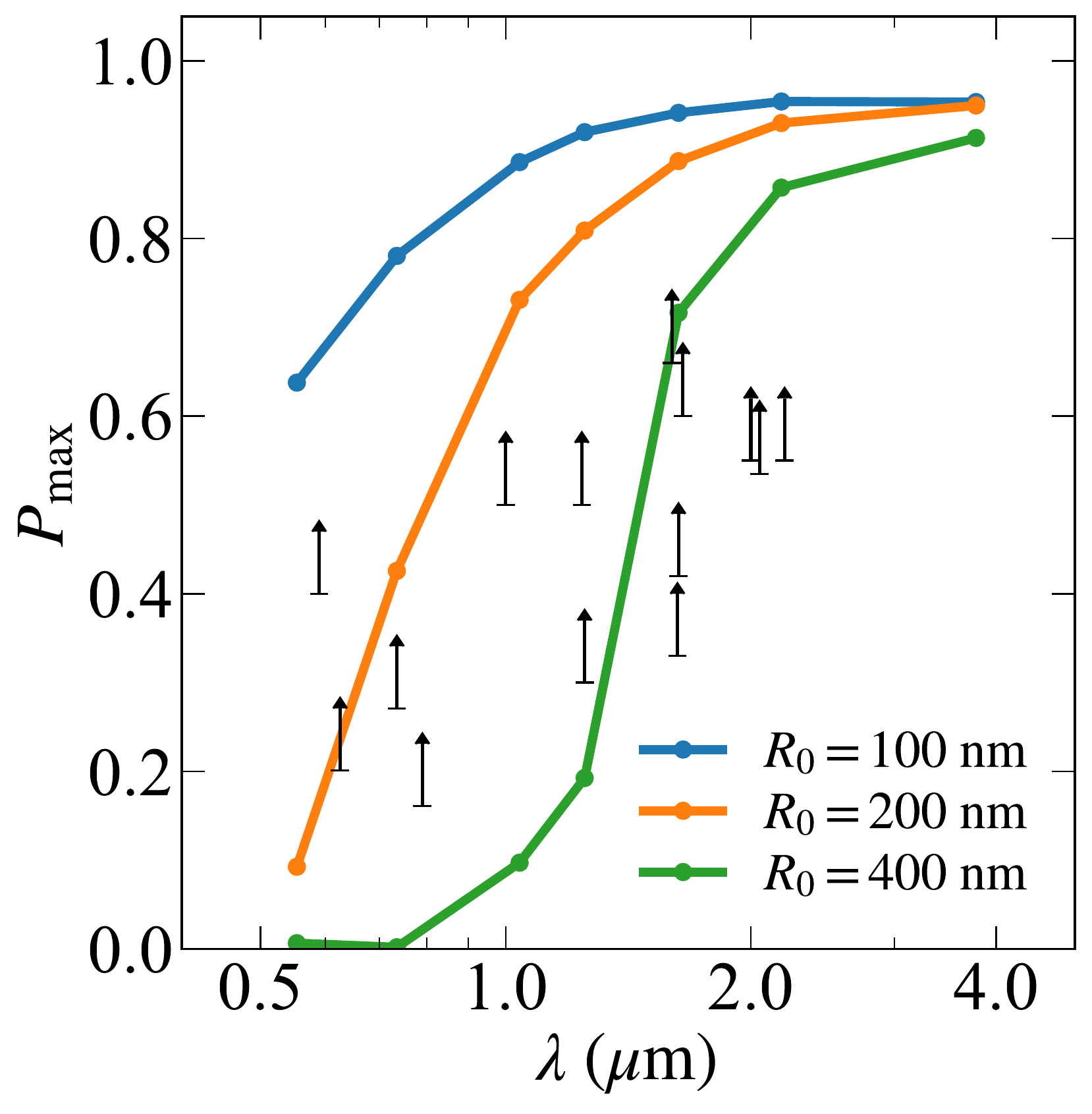}
\caption{Effect of monomer size for aggregates with a porosity of $\sim87\%$. The {\it upper panel} shows the degree of linear polarization of light scattered by aggregates with the \org~composition at $\lambda=1.04~\mu$m. The blue, orange, and green lines are the results for $\rmon=100$, $200$, $400$ nm, respectively. The {\it lower panel} shows the wavelength dependence of the maximum polarization. The observed maximum polarization fractions are plotted as lower limits, since the amount of depolarization that the observed polarized light suffered is unknown.}
\label{fig:pang}
\end{center}
\end{figure}

\subsection{The effect of monomer size on scattering polarization} \label{sec:mon}

When we attempt to derive the monomer radius from polarized disk scattered light, one of the fundamental questions is: {\it If the size, porosity, and composition (refractive index) of large aggregates ($\xc>1$) are the same, how does the degree of linear polarization depend on the monomer radius?}
To answer this, we first consider BPCA clusters with the \org~composition and $\rc=3.1~\mu$m, as shown in Fig. \ref{fig:agg} (upper panel).

To assess the effect of monomer radius on scattering polarization, we need to fix the parameters of aggregates other than the monomer radius, such as porosity and characteristic radius. Since BPCA clusters have a fractal dimension of 3, the characteristic radius scales as $N^{1/3}$. 
To fix $\rc$, the number of monomers should be decreased by a factor of 8 if the monomer radius is increased by a factor of 2. Therefore, we consider three aggregates having the following sets of parameters ($N$, $\rmon$)=($4096$, 100 nm), ($512$, 200 nm), and ($64$, 400 nm). All of them have nearly the same characteristic radii and porosities ($\rc$, $\mathcal{P}$)=($3.14~\mu$m, $86.8\%$), ($3.12~\mu$m, $86.4\%$), and ($3.12~\mu$m, $86.5\%)$ for $\rmon=100$, $200$, and $400$ nm, respectively. Furthermore, the three aggregates have the same material volume, as they have $\rv=\rmon N^{1/3}=1.6~\mu$m.

Fig. \ref{fig:pang} shows the degree of linear polarization of scattered light for unpolarized incoming light ($-S_{12}/S_{11}$) for the three aggregates, where $S_{ij}$ represents a scattering matrix element \citep{Bohren83}. Since we have fixed the parameters other than the monomer radius, the difference in the polarization can be attributed to a monomer-size effect.

Fig. \ref{fig:pang} (upper panel) shows the angular dependence of polarization at $\lambda=1.04~\mu$m. At this wavelength, the size parameters of the three aggregates are $\xc\simeq19$, and the size parameters of the monomers are $\xmon=0.6$, $1.2$, and $2.4$ for $\rmon=100$, $200$, and $400$ nm, respectively. 
For $\xmon=0.6$ ($\rmon=100$ nm), the polarization curve appears to be very similar to that of Rayleigh scattering despite the large aggregate size ($\xc\gg1$). In contrast, for $\xmon>1$ ($\rmon=200$ and $400$ nm), the degree of polarization drops rapidly and eventually oscillates around zero. In this way, the monomer radius has a significant impact on the degree of polarization of aggregates.

Fig. \ref{fig:pang} (lower panel) shows the wavelength dependence of the maximum degree of polarization $\pmax$, which is defined as the maximum value of $-S_{12}/S_{11}$ for all scattering angles.
For the case of Fig. \ref{fig:pang} (upper panel), $\pmax$ are $89\%$, 73\%, 9.7\% for $\rmon=100$ nm, 200 nm, and 400 nm, respectively. 
The maximum polarization is found to be almost monotonically decreasing with decreasing $\lambda$ for all cases. At $\lambda=3.78~\mu$m, the size parameters of the three monomers are all lie in $\xmon<0.67$. In this case, the difference in $\pmax$ between different monomer sizes is less than $5\%$ points, and thus, changing the monomer radius does not result in changing the maximum polarization.
The polarization for $\rmon=400$ nm starts to decrease at $\lambda=2.18~\mu$m, which corresponds to $\xmon\simeq1.2$. The difference in $\pmax$ between different monomer sizes now reaches $\sim 10\%$ points, and becomes more prominent for a shorter wavelength. Therefore, the effect of monomer size seems to appear only when the monomer is sufficiently large, i.e., $\xmon\gtrsim1$. 

Fig. \ref{fig:pang} suggests that the aggregates consisting of 400-nm monomers fail to explain a high observed polarization fraction at $\lambda\lesssim1.25~\mu$m for several disks. Although a more detailed parameter study and the comparison with observations will be presented later (see Sects. \ref{sec:param} and \ref{sec:obs}), this result already hints at the presence of relatively small monomers in disks.

\begin{figure}[t]
\begin{center}
\includegraphics[width=0.96\linewidth]{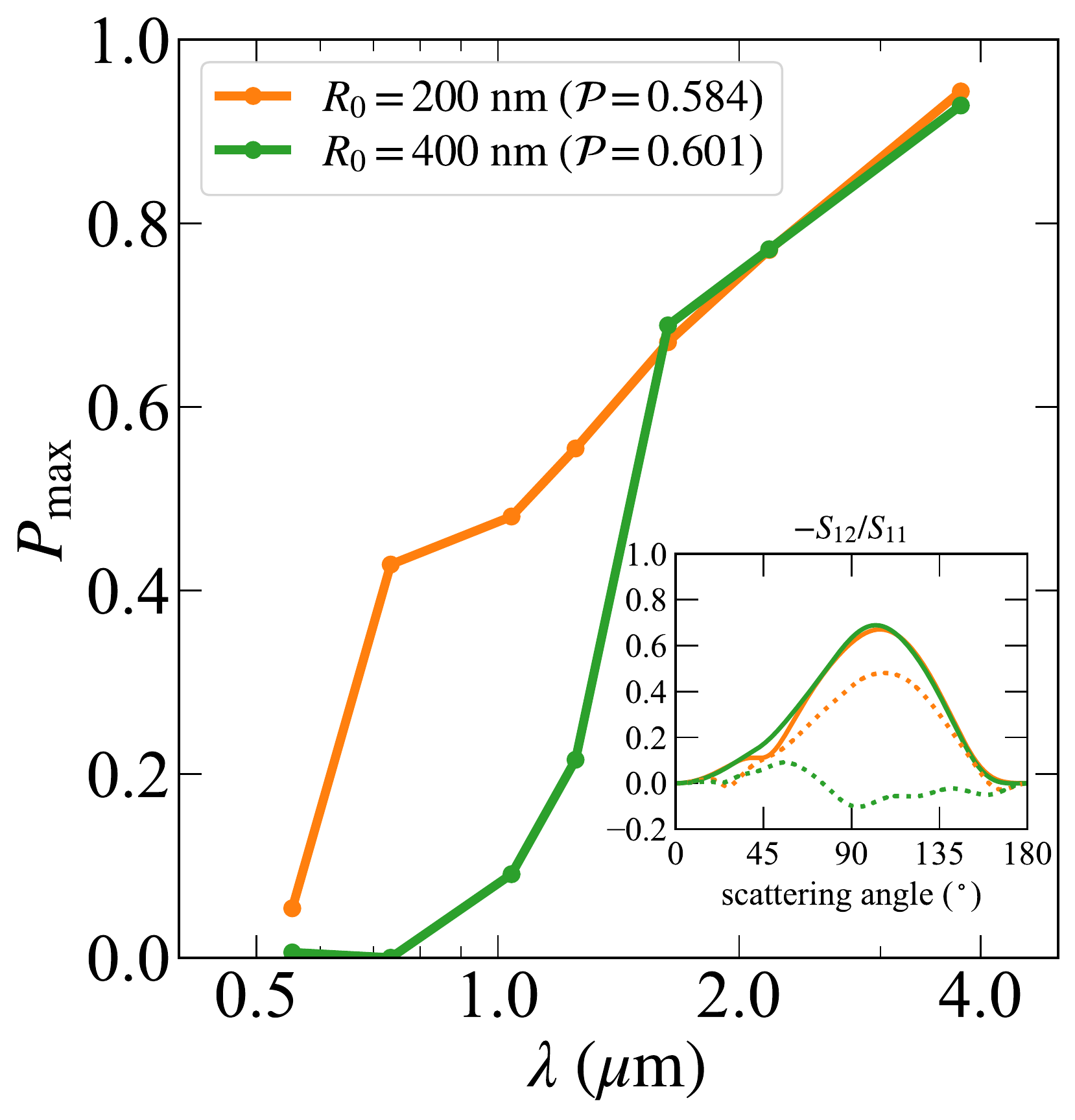}
\caption{Effect of monomer size for the aggregates with a porosity of $\sim59\%$. The orange and green lines are the results for $\rmon=200$ nm (BAM2 clusters of 128 monomers) and 400 nm (BAM1 clusters of 16 monomers), respectively. The porosity of each aggregate is indicated in the legend. The inset panel shows the angular dependence of the degree of polarization ($-S_{12}/S_{11}$) at two wavelengths (solid lines: $\lambda=1.63~\mu$m, dotted lines: $\lambda=1.04~\mu$m).}
\label{fig:lowpor}
\end{center}
\end{figure}

\citet{Shen09} claimed that the monomer size has negligible effect on polarization for moderately porous aggregates ($\mathcal{P}\approx60\%$), although they only considered relatively small monomers ($\xmon\lesssim1.6$).
To compare our results with theirs, we selected two aggregates with a porosity of $\sim60\%$ as follows. The first selected aggregate is BAM2 clusters with 128 monomers with a radius of 200 nm. The other aggregate is BAM1 clusters with 16 monomers with a radius of 400 nm. The two aggregates have the same material volume ($\rv=\rmon N^{1/3}\simeq1.01~\mu$m). Their characteristic radii and porosities are also very similar: ($\rc$,$\mathcal{P}$)=($1.35~\mu$m, $58.4\%$) and ($1.37~\mu$m, $60.1\%$). Although there is still a subtle difference in porosity, we think that such a tiny difference does not significantly alter our conclusions.

Fig. \ref{fig:lowpor} shows the degree of polarization for the two low-porosity aggregates ($\sim59\%$).
We found that the wavelength dependence of $\pmax$ is surprisingly similar between the two when $\lambda\gtrsim1.63~\mu$m, which corresponds to $\xmon\lesssim 1.54$. Thus, it appears to reaffirm the conclusion made by \citet{Shen09}. However, we also found a significant difference in $\pmax$ at $\lambda\lesssim 1.25~\mu$m, which corresponds to $\xmon\gtrsim2.01$. Therefore, a manifestation of the monomer size effect needs a slightly larger $\xmon$ value for low-porosity aggregates, that is, $\xmon\gtrsim2$.
In other words, the absence of the monomer size effect in \citet{Shen09} is likely due to its limited range of size parameters ($\xmon\lesssim1.6$).

To summarize, even if we fix the aggregate size, porosity, and composition, the degree of polarization depends sensitively on the monomer size unless $\xmon\lesssim1$--$2$. This tendency has been confirmed for a wide range of aggregate porosity ($59$--$87\%$). Conversely, if aggregates consist of small monomers such that $\xmon\lesssim 1$, the polarization is insensitive to the monomer radius and is determined by the porosity, size, and composition of aggregates \citep[Fig. \ref{fig:lowpor}, see also][]{Shen09}.
This is in agreement with the findings from laboratory experiments \citep{Zerull93, Gustafson99}. Our results suggest that, if we try to distinguish between sub-micron- and micron-sized monomers, the optical and near-IR wavelengths are the optimal wavelengths, because the monomer size parameter is close to or larger than unity at these wavelengths. 

\subsection{The maximum polarization for various aggregates} \label{sec:param}
\begin{figure*}[t]
\begin{center}
\includegraphics[width=0.99\linewidth]{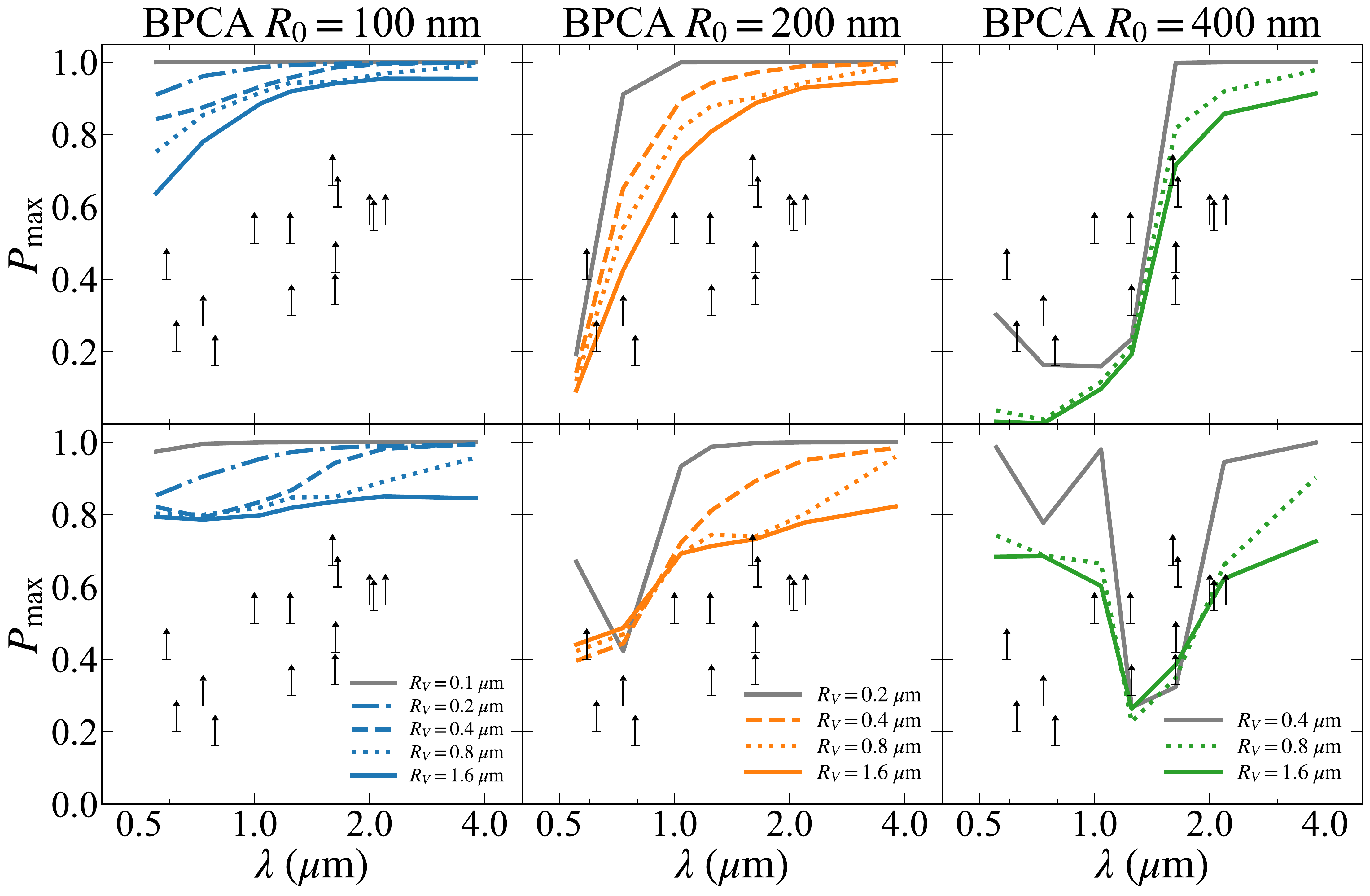}
\caption{Wavelength dependence of $\pmax$ for BPCA with the \org~ ({\it upper panels}) and \amc~ ({\it lower panels}) compositions. {\it From left to right}: the monomer radius of the aggregates is $\rmon=100$, $200$, and $400$ nm, respectively. The gray solid line in each panel represents $\pmax$ for a single spherical monomer. }
\label{fig:bpcaall}
\end{center}
\end{figure*}

We showed that the monomer size is important for the polarization characteristics of aggregates. Next, we study how the maximum polarization $\pmax$ depends on aggregate radius, porosity, composition, and monomer radius.

Fig. \ref{fig:bpcaall} shows the effect of aggregate radius on the maximum polarization for the case of BPCA. A similar plot for BAM1 and BAM2 is available in Appendix \ref{sec:appsize}. Apart from the porosity effect on the maximum polarization, the overall dependencies for BAM1 and BAM2 are similar to the case of BPCA.
Fig. \ref{fig:por} shows the effect of porosity on the maximum polarization. In those figures, we also plot the observed maximum polarization as a lower limit because the amount of depolarization that the observed polarized light suffered is unknown.
To estimate $\pmax$ values from $P_\mathrm{max}^\mathrm{obs}$, we need a more detailed modeling, i.e., radiative transfer calculations for each disk, which is beyond the scope of this paper.

The key dependencies seen in Figs. \ref{fig:bpcaall} and \ref{fig:por} are summarized below.

\subsubsection*{Dependence on aggregate radius}
Each panel of Fig. \ref{fig:bpcaall} shows the wavelength dependence of $\pmax$ for various aggregate radii. For comparison, we also plot the wavelength dependence for the single monomer (gray solid line). 
For \org-\texttt{100}, each monomer obeys Rayleigh scattering, and the maximum polarization is nearly 100\% at all wavelengths.
As with a single monomer, the maximum polarization of aggregates remains high at all wavelengths even if $\rv$ increases to $1.6~\mu$m (the characteristic radius of 3.1$~\mu$m). In contrast, for \org-\texttt{200} and -\texttt{400}, as $\pmax$ of the monomer drops, $\pmax$ of the aggregate does as well. The above results nicely illustrate the fact that the polarization for aggregates tends to reflect the polarization characteristics of the monomers, as known since the early 1990s \citep{West91, Kozasa93}. 

The resemblance of the polarization between aggregates and monomers is physically reasonable as multiple scattering is often subdominant (not negligible though) for highly porous aggregates ($\xc>1$ and $\xmon\ll1$) \citep{Berry86, Botet97, Tazaki16, Tazaki18}. In the single scattering limit, every nonzero scattering matrix element is equally affected due to the constructive and destructive interference of light waves from aggregates. Consequently, each scattering matrix element does depend on the aggregate structure, but the ratio of them, i.e., the degree of linear polarization, is independent of the aggregate structure and size and is solely determined by the properties of monomers \citep[see Eq. (9) in][]{Tazaki16}.

The deviation of $\pmax$ of the aggregates from that of the monomer is due to multiple scattering within each aggregate. Larger aggregates provide a higher probability of multiple scattering, resulting in lower $\pmax$ values.
However, the dependence on the aggregate radius is generally weak compared to that of the monomer radius. 
For example, even if we increase $\rv$ by more than an order of magnitude (from $0.1~\mu$m to $1.6~\mu$m), the reduction of $\pmax$ is less than a factor of two, whereas increasing the monomer radius only by a factor of two considerably affects the maximum polarization (Fig. \ref{fig:pang}). Such weak aggregate size dependencies have also been seen in previous numerical simulations \citep{Kozasa93, Kimura01, Kimura03, Kimura06, Petrova04, Bertini07} and micro-wave analog experiments \citep{Zerull93, Gustafson99}. 

\subsubsection*{Dependence on monomer composition}
Top and bottom panels of Fig. \ref{fig:bpcaall} show the maximum polarization for aggregates made of the \org~and \amc~composition, respectively. 
In general, the monomer composition has a strong impact on the maximum polarization. 

For all monomer radii, the maximum polarizations of the \amc~model are lower and higher than those of the \org~model at near-IR and optical wavelengths, respectively. Higher $n$ values or lower $k$ values tend to decrease the maximum polarization of aggregates \citep{Petrova04, Kimura06, Lumme11}, because such a change in refractive index tends to induce more multiple scattering within each aggregate. As shown in Table \ref{tab:opcont}, the \amc~model has a higher refractive index in both the real and imaginary parts than the \org~model. Because of the higher real part, $\pmax$ is smaller at near-infrared wavelengths, and because of the higher imaginary part, $\pmax$ is larger at optical wavelengths. As a result, the wavelength dependence of the maximum polarization of the \amc~model is shallower than that of the \org~model.

For $\rmon=400$ nm, the wavelength dependence is complex. For both compositions, the maximum polarization drops rapidly at $\lambda\sim1-2~\mu$m. This drop in polarization is triggered by a rapid reduction in the degree of polarization of the monomer itself (see gray line), not by multiple scattering within the aggregate.
We initially suspected that such a sharp drop is an artifact arising from the assumption of perfectly spherical monomers in our aggregate models and might be mitigated by considering more realistic, nonspherical monomers. However, comparing the degree of polarization of spherical and nonspherical particles, we found that a similar sharp drop can occur for nonspherical particles as well (see Appendix \ref{sec:shape} for more detailed discussion). Thus, we suppose that the sharp drop is a realistic property, although a more detailed study on the polarization of aggregates of nonspherical monomers is desirable.

\subsubsection*{Dependence on aggregate porosity}
\begin{figure*}[t]
\begin{center}
\includegraphics[width=0.99\linewidth]{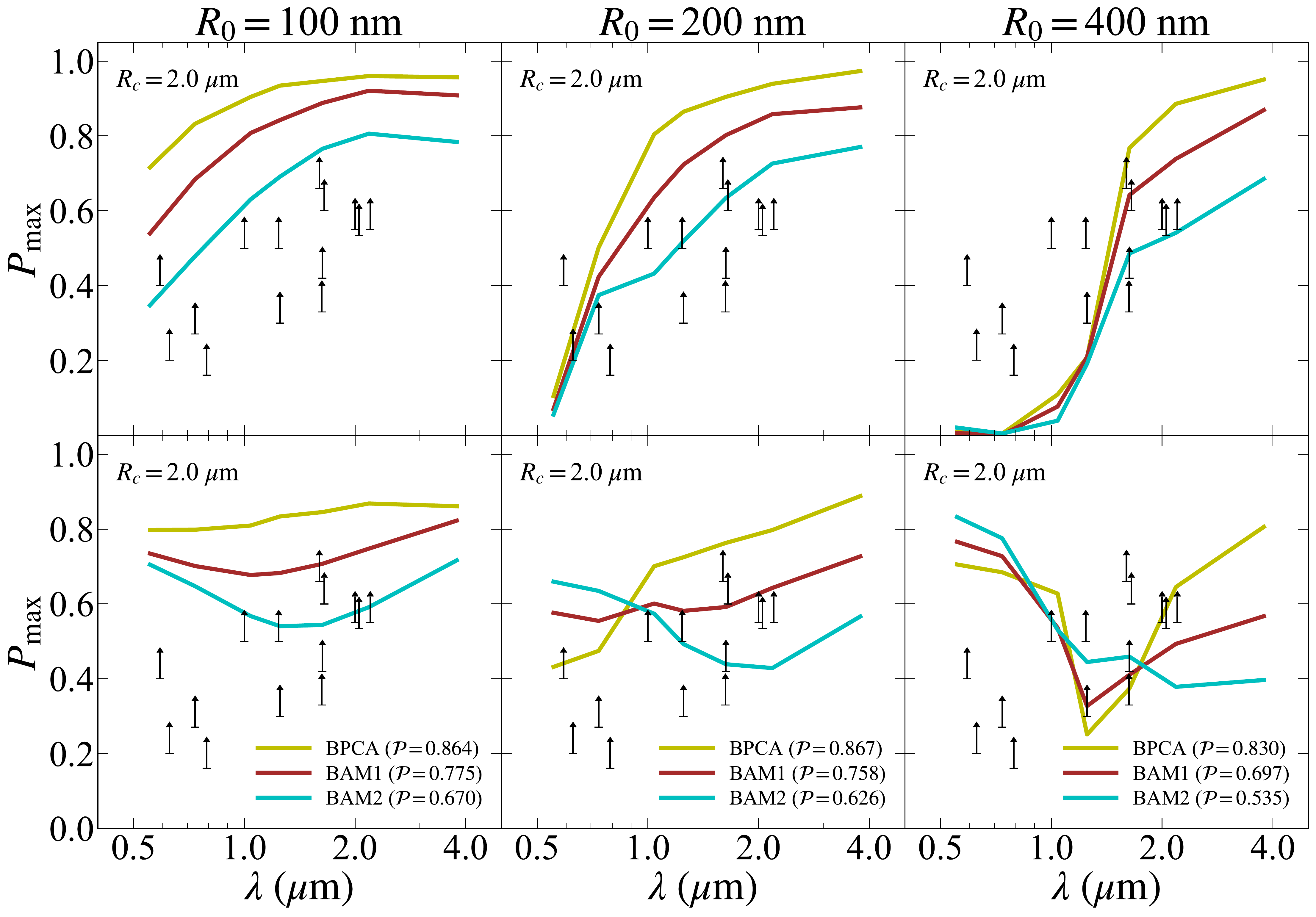}
\caption{Effect of porosity on the maximum polarization for the aggregates with $\rc=2.0~\mu$m with the \org~ ({\it upper panels}) and \amc~ ({\it lower panels}) compositions. {\it From left to right}: the monomer radius of the aggregates is $\rmon=100$, $200$, and $400$ nm, respectively. The yellow, brown, and cyan lines represent the results for BPCA, BAM1, and BAM2. The porosity of each aggregate model, $\mathcal{P}$, is indicated in each legend. }
\label{fig:por}
\end{center}
\end{figure*}

Fig. \ref{fig:por} shows the effect of porosity on the maximum polarization. To isolate the effect of porosity, we fixed the characteristic radius of aggregates to $\rc=2.0~\mu$m by interpolating the simulation data. 

For the \org~case, the maximum polarization monotonically decreases with decreasing $\lambda$ for all porosities. Decreasing the porosity also results in decreasing the maximum polarization, because a denser aggregate results in more multiple scattering \citep{Kolokolova10}. For the \amc~case, lower porosity aggregates (BAM1 and BAM2) exhibit an inverted wavelength dependence of the polarization: increasing $\pmax$ with decreasing $\lambda$. \citet{Shen09} also found a similar inverted trend for BAM2. Because of this inverted trend, the maximum polarization of BAM2 will be similar to or even slightly higher than that of BPCA at optical wavelengths.

We interpret this inverted behavior as follows. The wavelength at which the inversion sets in found to be the wavelength at which the radiation field in the aggregate begins to be non-uniform. The non-uniform radiation field is due to attenuation of the incoming radiation at the surface of the aggregate. With decreasing wavelength from this, the successive orders of scattering get fewer (less multiple scattering), which results in increasing $\pmax$. The above interpretation is further augmented by the following points. We confirmed that the radiation field for \org~aggregates is almost uniform for all porosities and wavelengths we investigated, and therefore they are not eligible for inversion. Although BPCA clusters with \amc~ composition exhibit a non-uniform radiation field, they are less efficient multiple scatterers in the first place, and hence the inversion effect does not appear clearly.

Another intriguing feature seen in Fig. \ref{fig:por} is that 
the porosity dependence diminishes for shorter wavelengths for \org~aggregates. This effect is reminiscent of the neighborhood effect \citep{Kimura04, Kolokolova10}; the degree of polarization is characterized by a local structure of a length scale of $\sim\lambda/2\pi$ or so within an aggregate. Once the size parameter of the monomers exceeds around unity, the electromagnetic interactions between neighboring monomers, rather than the overall aggregate structure, characterize the degree of polarization.
In our cases, the porosity dependence diminishes for $\lambda\le0.735~\mu$m for $\rmon=200$ nm and $\le1.25~\mu$m for $\rmon=400$ nm, which translate into $\xmon\ge1.71$ and $2.01$, respectively. In other words, the diminished porosity dependence occurs when the size parameter of monomers $\xmon\gtrsim1.71-2.01$ for the \org~composition. 

\citet{Kolokolova10} point out that the neighborhood effect appears easily for more absorbing aggregates. We confirmed that the converging behavior already starts to appear for \amc-\texttt{100} at optical wavelengths. However, the convergent behavior is missing for \amc-\texttt{200} and -\texttt{400}. We speculate that this is perhaps due to  (i) artifacts arising from ripple patterns in polarization curves and/or (ii) different coordination number of monomers, although further clarification seems difficult within the limit of our simulation data. 
As discussed in Appendix \ref{sec:shape}, strong ripple patterns were confirmed in the polarization curves for \amc-\texttt{200} and -\texttt{400} at an optical wavelength, while it is absent for \org-\texttt{200} and -\texttt{400}. Such a ripple pattern would not be present for realistic monomers, as the monomers must not be mono-disperse and perfectly spherical particles, and hence, their poly-dispersity and non-sphericity will smear them out. More detailed light scattering simulations taking into account aggregates of nonspherical monomers is necessary to clarify this point.

\subsection{How large are the monomers in planet-forming disks?} \label{sec:obs}
In Sects. \ref{sec:mon} and \ref{sec:param}, we investigated the wavelength dependence of the maximum polarization for various aggregates and clarified the impact of the monomer size on the polarization. 
In particular, as we discussed in Sect. \ref{sec:mon}, the optical and near-IR wavelengths are the optimal wavelengths for distinguishing sub-micron- and micron-sized monomers because the size parameter of the monomers becomes close to or larger than unity at these wavelengths. Comparing the simulation results with the observed maximum polarization fraction, i.e., Fig. \ref{fig:pobs}, we can assess the monomer radius of aggregates in planet-forming disks for the first time.
 
Currently available disk measurements favor the presence of relatively small monomers.
As shown in Figs. \ref{fig:bpcaall} and \ref{fig:por}, the aggregates with monomers of $\rmon=400$ nm fail to reproduce the combination of a high polarization fraction and a reddish polarization color at optical and near-IR wavelengths {\it regardless of the aggregate size, composition, and porosity}. For the nebula scattered light around HL Tau, the observed maximum polarization fractions may be marginally explained by aggregates with \amc-\texttt{400} unless the observed signals are depolarized by multiple scattering. This is in harmony with the findings of \citet{Murakawa08}. Using a one-dimensional single scattering model with the Mie theory, \citet{Murakawa08} modeled polarized scattered light from the HL Tau nebulosity and then estimated the maximum grain radius to be $0.4~\mu$m. Since we have shown that the degree of polarization of aggregates reflects the monomer property, the maximum grain radius they derived may be interpreted as the monomer radius rather than the aggregate size. 

Monomers larger than $\rmon=400$ nm would be unfavorable because their polarization will be even lower and/or bluish at an optical wavelength. As a result, we arrive at the conclusion that the monomers of aggregates in planet-forming disks are likely no greater than $400$ nm ($=0.4~\mu$m). The inferred monomer size is similar to the one seen in cometary dust aggregates and the maximum size of interstellar grains (see Sect. \ref{sec:dismon}). 

Although a monomer radius of 100 or 200 nm is consistent with the observations, we cannot exclude the presence of monomers much smaller than 100 nm.
This point will be clarified by studying the presence/absence of the variation in $P_\mathrm{max}^\mathrm{obs}$ at optical wavelengths across various disks. 
In Fig. \ref{fig:por}, we showed that the maximum polarization is insensitive to aggregate porosity when $\xmon\gtrsim1.71$--$2.01$ at least for the \org~composition.
This may imply that if aggregate porosity is diverse from one disk to another, the near-IR maximum polarization fractions should vary accordingly, but such a variation is suppressed for optical wavelengths when $\rmon>100$ nm. Conversely, if the variation at the optical wavelength is similar to the near-IR counterpart, the monomer radius is likely less than 100 nm. 
This strategy of deriving the monomer radius has been employed in the cometary field \citep{Gustafson99, Kimura03, Kimura06}. 

Fig. \ref{fig:pobs} shows that the variation in optical and near-IR polarizations are around $\sim16\%$ points and $\sim30\%$ points, hinting at less variation for shorter wavelengths, and this may indicate the presence of monomers larger than 100 nm. However, the number of current observational data points is rather limited, and therefore the time is not yet ripe to draw a robust conclusion about the lower size limit. A future survey on optical polarization will be crucial to strengthen this argument. 

Fig. \ref{fig:pobs} also points to a red polarization color. Within the limits of our composition models, dust aggregates in disks are unlikely to be dark and dense aggregates, i.e., BAM2 with \amc~composition, because such aggregates tend to exhibit an inverted, or blueish, wavelength dependence. In other words, such an inverted trend is smoking gun evidence of dark and dense aggregates (or can be dark monolithic grains). However, it is worth keeping in mind that the wavelength dependence could be affected by the assumption of a refractive index. If one allows more conducting materials (i.e., amorphous carbon or metals) or less dielectric materials (i.e., water ice), $|m|$ and $d|m|/d\lambda$ will be larger than our $\amc$ model. In this case, the inverted trend might be alleviated, although it is unclear whether such a composition model is reasonable for protoplanetary dust. In any case, the wavelength dependence of polarization is useful for diagnosing not only the monomer radius but also the monomer composition. 

\section{Application to the HD 142527 disk} \label{sec:142527}

Our argument so far has been based on the premise of the working hypothesis, that is, the monomer radius and its composition are the same for various disks. However, this may not be always true, and hence, each aggregate model does not necessarily satisfy all lower limits shown in Figs. \ref{fig:bpcaall} and \ref{fig:por}. Therefore, multiwavelength quantitative polarimetry for a specific planet-forming disk is crucial for a monomer size study. Here, as a case study, we focus on the disk around HD 142527, because polarization measurements have been made with the widest wavelength coverage among the disks shown in Fig. \ref{fig:pobs}.
 
HD 142527 is a binary star surrounded by a large circumbinary disk, whose optical/near-IR scattered light has been extensively investigated \citep{Fukagawa06, Rameau12, Avenhaus14, Avenhaus17, Rodigas14, Hunziker21}. \citet{Hunziker21} found that a high polarization fraction toward this disk, hinting at aggregate nature of dust particles. \citet{Tazaki21} found that the observed disk color is consistent with dust particles of a radius of $\sim3~\mu$m. 
Also, the reddish scattered light is more likely explained by compact aggregates\footnote{The meaning of `compact aggregates' here is those having a higher fractal dimension \citep[see Fig. 10 in ][]{Tazaki19}. Thus, this statement does not necessarily mean low porosity aggregates. For example, BPCA clusters have a fractal dimension of 3 and are thought to be eligible for producing a reddish color, but their porosity of $\sim85\%$ is still a class of highly porous.} rather than fractal aggregates \citep{Min16, Tazaki19}. From these results, compact dust aggregates of $\sim3~\mu\mathrm{m}$ in radius are thought to be responsible for the scattered light of the disk. Such aggregates are quite similar to those presented in Fig. \ref{fig:agg}, and therefore we may apply our simulation results to interpret the observations.

\citet{Hunziker21} conducted a quantitative polarimetric measurement of the disk around HD 142527 at optical (the VBB) and near-IR (the $H$-band) wavelengths. They accurately derived disk polarization fractions at the far side: $28.0\pm0.9\%$ at the VBB and $35.1\pm2.1\%$ at the $H$-band. They also argued that the maximum polarization fraction may be as high as $30\pm5\%$ for the VBB and $40\pm10\%$ for the $H$-band. 
In this study, we adopt $28.0\pm0.9\%$ and $35.1\pm2.1\%$ for the maximum polarization fractions at optical and near-IR wavelengths for this disk because the disk far side is close to a scattering angle of 90 degrees at which the degree of polarization is often maximized (see Fig. \ref{fig:pang}) and better accuracy of the data.

Since the observed polarization fractions suffer a depolarization effect due to multiple scattering at the disk surface, we converted the observed maximum polarization fraction to $\pmax$ by using a simple plane-parallel model developed by \citet{Ma22}. The depolarization effect depends on the single scattering albedo of dust aggregates. The single scattering albedo of our aggregates with $\rv=1.6~\mu$m ranges from 0.5--0.95. With this range, the plane-parallel model predicts that $\pmax\sim42-86\%$ for the VBB and $51-100\%$ for the $H$-band. The estimated range agrees with the $\pmax$ values estimated in \citet{Hunziker21}, where they derived $\pmax=50~\%$ and $70~\%$ at VBB and the $H$-band, respectively.

\begin{figure}[t]
\begin{center}
\includegraphics[width=\linewidth]{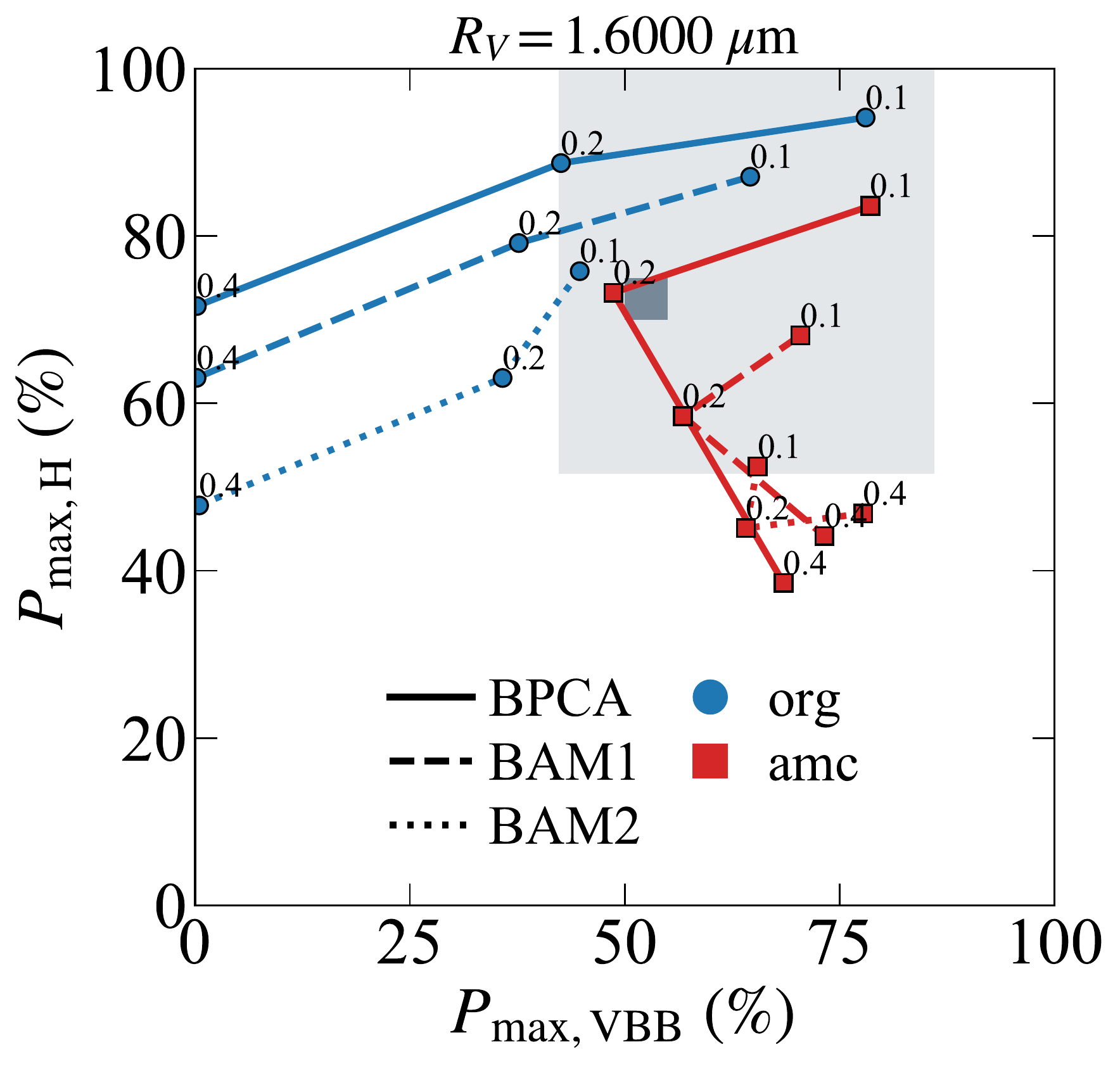}
\caption{Maximum polarization of dust aggregates at the VBB (optical) and the $H$-band (near-IR) with $\rv=1.6~\mu$m. Different line styles correspond to different aggregate models (BPCA: solid, BAM1: dashed, BAM2: dotted lines), and different symbols correspond to different composition models (circles: \org~, squares: \amc). The number beside each point denotes the monomer radius in unit of $\mu$m. The dark gray hatched area is the maximum polarization estimated by \citet{Hunziker21}, while the light gray hatched area is the estimated $\pmax$ values for an assumed range of the single scattering albedo of $0.5$--$0.95$.}
\label{fig:hd142527}
\end{center}
\end{figure}

Fig. \ref{fig:hd142527} summarizes $\pmax$ values at the VBB and the $H$-band for various aggregate porosities, monomer radii, and compositions. We assumed the aggregate radius of $\rv=1.6~\mu$m, and then characteristic radii are around $2.3$--$3.1~\mu$m, which approximately agree with the sizes inferred from a previous modeling \citep{Tazaki21}. The inferred observational range of $\pmax$ is shown with the gray-hatched square in the plot. The inferred range falls onto the region where $\rmon=100$--$200$ nm, while aggregates with $\rmon=400$ nm produce either too small optical or near-IR polarization fraction. One may wonder if an intermediate composition between \amc~and \org~yields a closer fit to the observations. However, we found that even if we consider an intermediate composition, the large monomer still fails to reproduce the estimated range of maximum polarization (Appendix \ref{sec:intermed}).  

\begin{figure}[t]
\begin{center}
\includegraphics[width=\linewidth]{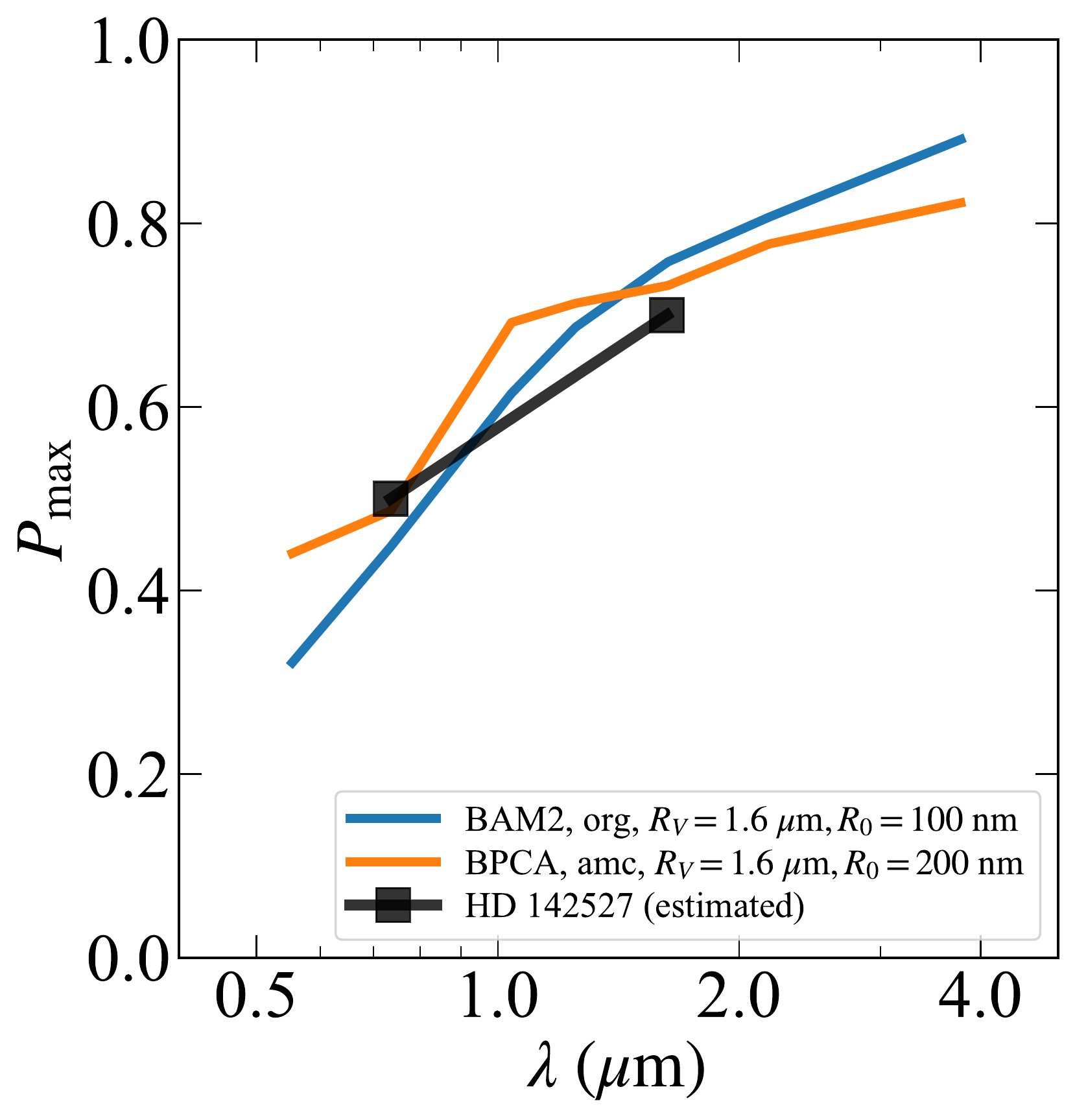}
\caption{Wavelength dependence of the maximum polarization toward the disk around HD 142527. The orange and blue lines show BPCA clusters with \amc-\texttt{200} monomers and BAM2 clusters with \org-\texttt{100} monomers, respectively. Both aggregate models have $\rv=1.6~\mu$m.}
\label{fig:best}
\end{center}
\end{figure}

Therefore, the wavelength dependence of the maximum polarization fractions of the HD 142527 disk is consistent with large aggregates ($\sim3~\mu$m) with relatively small monomers ($\rmon=100$--$200$ nm). In particular, the $\pmax$ values estimated by \citet{Hunziker21} are similar to the model of BPCA clusters with \amc--\texttt{200} monomers or BAM2 clusters with \org--\texttt{100} monomers (Fig. \ref{fig:best}). To distinguish between the two possibilities, a further detailed study is necessary, which is left for our future tasks.

\section{Discussion} \label{sec:discussion}

\subsection{Comparison with previous estimates on the monomer radius} \label{sec:dismon}

We have demonstrated that the monomer radius is likely no greater than 400 nm to explain the observed polarization fractions. We are aware, however, that, 
our results may be biased toward aggregates with smaller monomers because aggregates with larger monomers, if present, may sink below the disk scattering surface probed by optical and near-IR observations. This issue is inevitable as the younger disks are generally optically thick at shorter wavelengths. 

One possibility to overcome this issue is to observe debris disks, which are optically thin.
\citet{Graham07} showed that scattered light around AU Mic may be explained by fluffy, presumably primordial, aggregates. At a wavelength of $\lambda=0.590~\mu$m, the observed polarization fraction is as high as $0.4$, and the authors estimated the maximum polarization of each aggregate to be $\pmax>0.5$. Although a high polarization at an optical wavelength could be explained by large and dark compact aggregates (see Fig. \ref{fig:por} {\it lower, right panel}), these aggregates tend to make the disk color reddish \citep{Min16}, while the fluffy aggregates makes it gray or blueish \citep{Tazaki19}. Given bluish scattered light around AU Mic \citep{Fitzgerald07}, fluffy aggregates with sub-micron monomers seems favorable. A similarity of the monomer size between optically thick younger disks and a debris disk lends further credibility to that sub-micron-sized monomers are a good representative of a dominant building block of aggregates. 

It is worth mentioning that there are some debris disks that show qualitatively different scattering polarization properties from the AU Mic disk, such as HR 4796 A. For HR 4796 A, the total intensity phase function exhibits enhanced back scattering \citep{Milli17}, and the degree of polarization peaks at a scattering angle of $40$--$50$ degrees \citep{Perrin15, Arriaga20}.
None of our simulation results satisfies these properties. However, very large ($\gtrsim 100~\mu\mathrm{m}$), densely packed aggregates of sub-micron- and micron-sized monomers may explain such observations \citep{Markkanen18}. Thus, light scattering simulations for very large aggregates would be crucial to constrain the monomer sizes for such disks.

The monomer sizes inferred in this study are smaller than those needed to explain the disk observations of millimeter-wave scattering polarization ($\rmon\gtrsim\mu$m) \citep{Okuzumi19, Arakawa21} (see Sect. \ref{sec:ice}). On the other hand, the inferred sizes are reminiscent of the maximum sizes of interstellar grains or typical subunits sizes seen in solar system dust aggregates. For the interstellar grain model, the maximum grain radius of $250$ nm has been often employed \citep{MRN77, Draine84}. A more recent model also adopts a grain size distribution peaked at $\sim200$ nm \citep{Jones13}. 

In the solar system, the chondritic porous interplanetary dust particle (CP IDP) typically exhibits a cluster-of-grapes morphology with an average subunit size of 300 nm in diameter \citep{Brownlee85}. \citet{Rietmeijer93} measured the subunit-size distribution of one CP IDP and found that the subunit size spans from 64 to 7580 nm with a mean of 585 nm in diameter and follows a log-normal distribution. One of the major constituents of CP IDP is GEMS (glass with embedded metal and sulfide). \citet{Woz13} measured the size distribution of GEMS in four CP IDPs. The geometric mean subunit sizes are ranging from 76--138 nm in radius for the four CP IDPs. Despite its similarity to the interstellar grain sizes, GEMS has been considered to be the condensates at the early solar nebula rather than the one directly inherited from the interstellar medium \citep{Keller11}. 

Recent in situ measurements of dust aggregates of the comet 67P/C-G by the Rosetta/MIDAS instrument have also revealed the presence of sub-micron-sized subunits. \citet{Bentley16} identified seven subunits in particle D ($\sim1.09^{+0.01}_{-0.25}~\mu$m in size), using a 80-nm-scan resolution, which are ranging from 
$0.260^{+0.050}_{-0.120}$ $\mu$m to $0.540^{+0.020}_{-0.250}$ $\mu$m in diameter.  \citet{Mannel16} also derived a subunit distribution of larger particles E and F and obtained the mean subunit diameter of $1.48^{+0.13}_{-0.59}~\mu$m and $1.36^{+0.15}_{-0.59}~\mu$m, respectively, although the scan resolutions of particles E and F are 210 nm and 195 nm, respectively, and thereby are worse than those used to analyze particle D. \citet{Mannel19} analyzed the subunits of particle G with a high-resolution scan (8 nm resolution) and derived the arithmetic mean subunit diameter of $99.49^{+0.89}_{-6.41}$ nm. They also hypothesized that particles D, E, and F, which were analyzed with a lower scan resolution in earlier studies, also consist of $\sim100$ nm subunits in diameter. 

The tensile strength of the comet 67P/C-G provides another way to study the cometary monomer size. Based on a formula derived from numerical simulations, \citet{Tatsuuma19} estimated a monomer radius to be $3.3$--$220~\mu\mathrm{m}$ to explain the tensile strengths. \citet{K20tensile} claims, however, that sub-micron sized monomers can still explain the observation if the volume effect in tensile strengths plays a role. In this way, the monomer radius derived from the tensile strength of the comet 67P/C-G seems still inconclusive and further study is necessary.

Although our results do not allow to distinguish the origin of monomers, such as interstellar origin and condensation at the disk forming epoch, the common statement that the planet formation begins with coagulation of sub-micron monomers has been confirmed from the vantage of disk observations for the first time. 

\subsection{Are icy aggregates in planet-forming disks sticky?} \label{sec:ice}

Recent detections of millimeter-wave scattering polarization provide new insight into the aggregate sizes in planet-forming disks \citep{Kataoka16, Stephens17, Hull18, Ohashi18, Bacciotti18, Dent19, Ohashi19, Ohashi20, Mori21}. To reproduce the millimeter-wave polarization of the HL Tau disk, \citet{Okuzumi19} argued that the critical fragmentation velocity should be as low as $0.1-1$ m s$^{-1}$. However, the origin of such a low critical fragmentation velocity is still unknown.

The critical fragmentation velocity is known to scale as \citep{Dominik97} 
\begin{equation}
u_\mathrm{f}=u_\mathrm{f0}\left(\frac{\rmon}{100~\mathrm{nm}}\right)^{-5/6}, \label{eq:vfrag}
\end{equation}
where $u_\mathrm{f0}$ is the fragmentation velocity of aggregates consist of 0.1-$\mu$m-sized monomers. In general, $u_\mathrm{f0}$ depends on surface materials coating each monomer and aggregate structure. To explain the suggested low $u_\mathrm{f}$ value, two possibilities exist (i) larger monomer radii ($\rmon\gtrsim\mu$m) and/or (ii) lower adhesion forces (a low $u_\mathrm{f0}$ value). As we found that the monomer radius in the HL Tau nebula may be around 0.4 $\mu$m (Sec. \ref{sec:obs}), the second possibility appears to be more likely. When $\rmon=400$ nm, we need $u_\mathrm{f0}\lesssim3~\mathrm{m~s}^{-1}$.

For H$_2$O-ice aggregates, by performing numerical collisional simulations, \citet{Wada09} derived $u_\mathrm{f0}=50~\mathrm{m~s}^{-1}$ for collisions between equal-sized aggregates. The velocity may vary from $u_\mathrm{f0}\sim 25$--$80$ m s$^{-1}$ depending on the mass ratio of colliding aggregates \citep{Wada13, Hasegawa21}. Laboratory experiments have also supported this highly sticky properties of H$_2$O ice \citep{Gundlach15}. Although recent experiments suggest a poorer stickiness of H$_2$O ice at a lower temperature \citep{Musiolik19, Gundlach18}, the interpretation of temperature dependence is still a matter of discussion \citep{K20ice, K20tensile}. If H$_2$O ice is as sticky as previously thought, the values of $u_\mathrm{f0}$ are an order of magnitude higher than our necessary value, $u_\mathrm{f0}\lesssim3~\mathrm{m~s}^{-1}$. Therefore, H$_2$O ice would not be the primary surface composition of monomers, at least in outer disk regions we can observe. 

One possible explanation is the presence of CO$_2$ ice on the monomer surface. Laboratory experiments have shown that CO$_2$ ice is much less sticky than H$_2$O ice \citep{Musiolik16a, Musiolik16b, Fritscher21}, i.e., $u_\mathrm{f0}\sim5~\mathrm{m~s}^{-1}$ \citep{Okuzumi19}. \citet{Arakawa21} attributed the higher stickiness of H$_2$O ice than CO$_2$ to viscoelastic dissipation upon collisions. Due to its poor stickiness, the presence of CO$_2$ ice has significant impact on dust coagulation outside the CO$_2$ snowline \citep{Pinilla17, Okuzumi19}. Although the value ($u_\mathrm{f0}\sim5~\mathrm{m~s}^{-1}$) is still higher than the necessary values, it has a similar order of magnitude. Thus, we speculate that monomers at the outer disk regions are covered by non-sticky ice, such as CO$_2$, and then efficient collisional fragmentation gives rise to relatively small aggregate sizes, as inferred from observations.

\subsection{Can coagulation trigger Streaming instability?}

One scenario for planetesimal formation is the streaming instability \citep{Youdin05} followed by a strong dust clumping \citep{Johansen07Nat, Johansen09, Bai10, Bai10pi, Sekiya18, Carrera21}. 
In the absence of external turbulence, dust settling will set the maximum dust-to-gas ratio, which is proportional to the metallicity. The threshold metallicity required to trigger the strong clumping rapidly increases with decreasing the particle Stokes number \citep{Carrera15, Yang17, Li21}.
One key question is whether collisional growth results in a Stokes number large enough to trigger the strong clumping \citep{Draz14, Lorek18}. 

If dust coagulation in the disks is limited by collisional fragmentation, the Stokes number of the maximum grain radius will be given by \citep{Birnstiel12}
\begin{equation}
\mathrm{St}_\mathrm{frag}=\frac{f_\mathrm{f}}{3}\frac{u_\mathrm{f}^2}{\alphat c_\mathrm{s}^2}, \label{eq:stfrag}
\end{equation}
where $\alphat$ is the turbulence parameter, $c_\mathrm{s}$ is the sound speed, and $f_\mathrm{f}=0.37$ is a numerical factor. It is important to notice the fact that $\mathrm{St}_\mathrm{frag}$ no longer depends on the gas surface density and the material density. By substituting Eq. (\ref{eq:vfrag}) into (\ref{eq:stfrag}), we obtain
\begin{equation}
\mathrm{St}_\mathrm{frag}\simeq0.02\left(\frac{u_\mathrm{f0}}{5~\mathrm{m~s}^{-1}}\right)^{2}\left(\frac{\rmon}{100~\mathrm{nm}}\right)^{-5/3}\left(\frac{\alphat}{10^{-3}}\right)^{-1},
\end{equation}
where we have used $c_\mathrm{s}=3.8\times10^4$ cm s$^{-1}$.

We estimate the metallicity $Z$ needed to trigger strong clumping by using a model by \citet{Li21}. Assuming a typical pressure gradient parameter $\Pi=0.05$, we obtain $Z\ge0.1$ for $\alpha=10^{-3}$ and $\mathrm{St}_\mathrm{frag}=0.02$, and  $Z\ge0.008$ for $\alpha=10^{-4}$ and $\mathrm{St}_\mathrm{frag}=0.2$. In the moderate turbulence case ($\alphat=10^{-3}$), particles should accumulate approximately 10 times with respect to the gas. 

Therefore, even in the outer region of the disk, efficient particle accumulation seems necessary to trigger strong clumping unless the turbulence is extremely low. In the above estimate, the role of turbulence is to just determine the thickness of the dust layer. However, a role for turbulence would not be that simple. \citet{Johansen07Nat} found that turbulence driven by magneto-rotational instability may create a temporary overdense region, which promotes the subsequent clumping by the streaming instability. In this way, on the one hand, external turbulence renders clumping challenging. On the other hand, it promotes clumping. Therefore, clarifying the interplay between external turbulence and the streaming instability would be essential to draw conclusions over the question if it works out in a turbulent disk.

\section{Summary} \label{sec:summary}

We have studied the effect of monomer size and composition on scattering polarization of dust aggregates by using an exact light scattering technique. Applying the simulation results to disk observations, we have estimated the monomer radius of aggregates in planet-forming disks for the first time. The main findings of this paper are as follows.

\begin{itemize}

\item The maximum degree of polarization of aggregates sensitively depends on the monomer size when the monomer size parameter $\xmon\gtrsim1$--$2$. Contrary to the previous study, we showed that the effect of monomer size appears not only for highly porous aggregates ($\mathcal{P}\sim87\%$) but also for less porous aggregates ($\mathcal{P}\sim59\%$), although less porous aggregates need a slightly larger monomer size for this effect to appear (Sect. \ref{sec:mon}).

\item In contrast, the maximum degree of polarization of aggregates becomes insensitive to monomer size when $\xmon\lesssim1$. In this case, the maximum polarization depends mainly on aggregate size, porosity, and composition. In general, aggregates with a larger size, lower porosity, or a higher real part of the refractive index tend to yield a lower maximum polarization fraction at near-IR wavelengths (Sects. \ref{sec:mon} and \ref{sec:param}).

\item Since the effect of monomer size is noticeable when $\xmon\gtrsim1-2$, the optical or near-IR wavelengths are the optimal wavelengths to distinguish between sub-micron-sized and micron-sized monomers because the monomer size parameter becomes close to or larger than unity at these wavelengths (Sect. \ref{sec:mon}). 

\item By comparing our results with the observations, we found that the monomer radius appears to be no greater than $0.4~\mu$m for several planet-forming disks (Sect. \ref{sec:obs}). We also found that a monomer radius of $0.1$--$0.2~\mu$m seems favorable to explain the recent polarimetric observations of the disk around HD 142527 (Sect. \ref{sec:142527}). 

\item It may be possible that the the monomers are much smaller than $0.1~\mu$m. In this case, our results predict that the variation in the maximum polarization fractions of various disks at optical wavelengths will be similar to that at near-IR wavelengths. In contrast, the variation in optical polarization would be small compared to the near-IR variation when the monomer are larger than 0.1 $\mu$m. A large optical polarization survey of planet-forming disks is therefore useful to test the monomer size (Sect. \ref{sec:obs}).

\item Within the limit of our composition model, we can rule out dark densely-packed aggregates at the disk surfaces, as they exhibit a bluish color of the maximum degree of polarization at optical and near-IR wavelengths, which is inconsistent with the observed reddish colors (Sect. \ref{sec:obs}). 
\end{itemize}

Optical and near-IR quantitative polarimetry will provide the observational grounds on the initial conditions for dust coagulation, and thereby, planetesimal formation in disks.

\begin{acknowledgements}
R.T. acknowledges the JSPS overseas research fellowship. We thank Daniel Mackowski and Maxim A. Yurkin for making the MSTM and ADDA codes publicly available, respectively. We also thank Bruce Draine for the availability of particle data of BA, BAM1, and BAM2. We thank J. Ma and H. M. Schmid for a fruitful discussion. 

\end{acknowledgements}

\bibliographystyle{aa} 
\bibliography{cite}

\begin{thebibliography}{131}
\expandafter\ifx\csname natexlab\endcsname\relax\def\natexlab#1{#1}\fi

\bibitem[{{Arakawa} \& {Krijt}(2021)}]{Arakawa21}
{Arakawa}, S. \& {Krijt}, S. 2021, \apj, 910, 130

\bibitem[{{Arriaga} {et~al.}(2020){Arriaga}, {Fitzgerald}, {Duch{\^e}ne},
  {Kalas}, {Millar-Blanchaer}, {Perrin}, {Chen}, {Mazoyer}, {Ammons}, {Bailey},
  {Barman}, {Bulger}, {Chilcote}, {Cotten}, {De Rosa}, {Doyon}, {Esposito},
  {Follette}, {Gerard}, {Goodsell}, {Graham}, {Greenbaum}, {Hibon}, {Hom},
  {Hung}, {Ingraham}, {Konopacky}, {Macintosh}, {Maire}, {Marchis}, {Marley},
  {Marois}, {Metchev}, {Nielsen}, {Oppenheimer}, {Palmer}, {Patience},
  {Poyneer}, {Pueyo}, {Rajan}, {Rameau}, {Rantakyr{\"o}}, {Ruffio},
  {Savransky}, {Schneider}, {Sivaramakrishnan}, {Song}, {Soummer}, {Thomas},
  {Wang}, {Ward-Duong}, \& {Wolff}}]{Arriaga20}
{Arriaga}, P., {Fitzgerald}, M.~P., {Duch{\^e}ne}, G., {et~al.} 2020, \aj, 160,
  79

\bibitem[{{Avenhaus} {et~al.}(2017){Avenhaus}, {Quanz}, {Schmid}, {Dominik},
  {Stolker}, {Ginski}, {de Boer}, {Szul{\'a}gyi}, {Garufi}, {Zurlo},
  {Hagelberg}, {Benisty}, {Henning}, {M{\'e}nard}, {Meyer}, {Baruffolo},
  {Bazzon}, {Beuzit}, {Costille}, {Dohlen}, {Girard}, {Gisler}, {Kasper},
  {Mouillet}, {Pragt}, {Roelfsema}, {Salasnich}, \& {Sauvage}}]{Avenhaus17}
{Avenhaus}, H., {Quanz}, S.~P., {Schmid}, H.~M., {et~al.} 2017, \aj, 154, 33

\bibitem[{{Avenhaus} {et~al.}(2014){Avenhaus}, {Quanz}, {Schmid}, {Meyer},
  {Garufi}, {Wolf}, \& {Dominik}}]{Avenhaus14}
{Avenhaus}, H., {Quanz}, S.~P., {Schmid}, H.~M., {et~al.} 2014, \apj, 781, 87

\bibitem[{{Bacciotti} {et~al.}(2018){Bacciotti}, {Girart}, {Padovani}, {Podio},
  {Paladino}, {Testi}, {Bianchi}, {Galli}, {Codella}, {Coffey}, {Favre}, \&
  {Fedele}}]{Bacciotti18}
{Bacciotti}, F., {Girart}, J.~M., {Padovani}, M., {et~al.} 2018, \apjl, 865,
  L12

\bibitem[{{Bai} \& {Stone}(2010{\natexlab{a}})}]{Bai10}
{Bai}, X.-N. \& {Stone}, J.~M. 2010{\natexlab{a}}, \apj, 722, 1437

\bibitem[{{Bai} \& {Stone}(2010{\natexlab{b}})}]{Bai10pi}
{Bai}, X.-N. \& {Stone}, J.~M. 2010{\natexlab{b}}, \apjl, 722, L220

\bibitem[{{Bentley} {et~al.}(2016){Bentley}, {Schmied}, {Mannel}, {Torkar},
  {Jeszenszky}, {Romstedt}, {Levasseur-Regourd}, {Weber}, {Jessberger},
  {Ehrenfreund}, {Koeberl}, \& {Havnes}}]{Bentley16}
{Bentley}, M.~S., {Schmied}, R., {Mannel}, T., {et~al.} 2016, \nat, 537, 73

\bibitem[{{Berry} \& {Percival}(1986)}]{Berry86}
{Berry}, M.~V. \& {Percival}, I.~C. 1986, Optica Acta, 33, 577

\bibitem[{{Bertini} {et~al.}(2007){Bertini}, {Thomas}, \&
  {Barbieri}}]{Bertini07}
{Bertini}, I., {Thomas}, N., \& {Barbieri}, C. 2007, \aap, 461, 351

\bibitem[{{Betti} {et~al.}(2022){Betti}, {Follette}, {Jorquera}, {Duch{\^e}ne},
  {Mazoyer}, {Bonnefoy}, {Chauvin}, {P{\'e}rez}, {Boccaletti}, {Pinte},
  {Weinberger}, {Grady}, {Close}, {Defr{\`e}re}, {Downey}, {Hinz},
  {M{\'e}nard}, {Schneider}, {Skemer}, \& {Vaz}}]{Betti22}
{Betti}, S.~K., {Follette}, K., {Jorquera}, S., {et~al.} 2022, arXiv e-prints,
  arXiv:2201.08868

\bibitem[{{Birnstiel} {et~al.}(2018){Birnstiel}, {Dullemond}, {Zhu}, {Andrews},
  {Bai}, {Wilner}, {Carpenter}, {Huang}, {Isella}, {Benisty}, {P{\'e}rez}, \&
  {Zhang}}]{Birnstiel18}
{Birnstiel}, T., {Dullemond}, C.~P., {Zhu}, Z., {et~al.} 2018, \apjl, 869, L45

\bibitem[{{Birnstiel} {et~al.}(2012){Birnstiel}, {Klahr}, \&
  {Ercolano}}]{Birnstiel12}
{Birnstiel}, T., {Klahr}, H., \& {Ercolano}, B. 2012, \aap, 539, A148

\bibitem[{{Bohren} \& {Huffman}(1983)}]{Bohren83}
{Bohren}, C.~F. \& {Huffman}, D.~R. 1983, {Absorption and scattering of light
  by small particles}

\bibitem[{{Botet} {et~al.}(1997){Botet}, {Rannou}, \& {Cabane}}]{Botet97}
{Botet}, R., {Rannou}, P., \& {Cabane}, M. 1997, \ao, 36, 8791

\bibitem[{{Brownlee}(1985)}]{Brownlee85}
{Brownlee}, D.~E. 1985, Annual Review of Earth and Planetary Sciences, 13, 147

\bibitem[{{Carrera} {et~al.}(2015){Carrera}, {Johansen}, \&
  {Davies}}]{Carrera15}
{Carrera}, D., {Johansen}, A., \& {Davies}, M.~B. 2015, \aap, 579, A43

\bibitem[{{Carrera} {et~al.}(2021){Carrera}, {Simon}, {Li}, {Kretke}, \&
  {Klahr}}]{Carrera21}
{Carrera}, D., {Simon}, J.~B., {Li}, R., {Kretke}, K.~A., \& {Klahr}, H. 2021,
  \aj, 161, 96

\bibitem[{{Chokshi} {et~al.}(1993){Chokshi}, {Tielens}, \&
  {Hollenbach}}]{Chokshi93}
{Chokshi}, A., {Tielens}, A.~G.~G.~M., \& {Hollenbach}, D. 1993, \apj, 407, 806

\bibitem[{{D'Alessio} {et~al.}(2001){D'Alessio}, {Calvet}, \&
  {Hartmann}}]{dalessio01}
{D'Alessio}, P., {Calvet}, N., \& {Hartmann}, L. 2001, \apj, 553, 321

\bibitem[{{Dent} {et~al.}(2019){Dent}, {Pinte}, {Cortes}, {M{\'e}nard},
  {Hales}, {Fomalont}, \& {de Gregorio-Monsalvo}}]{Dent19}
{Dent}, W.~R.~F., {Pinte}, C., {Cortes}, P.~C., {et~al.} 2019, \mnras, 482, L29

\bibitem[{{Dominik} \& {Tielens}(1997)}]{Dominik97}
{Dominik}, C. \& {Tielens}, A.~G.~G.~M. 1997, \apj, 480, 647

\bibitem[{{Dorschner} {et~al.}(1995){Dorschner}, {Begemann}, {Henning},
  {Jaeger}, \& {Mutschke}}]{Dorschner95}
{Dorschner}, J., {Begemann}, B., {Henning}, T., {Jaeger}, C., \& {Mutschke}, H.
  1995, \aap, 300, 503

\bibitem[{{Draine} \& {Lee}(1984)}]{Draine84}
{Draine}, B.~T. \& {Lee}, H.~M. 1984, \apj, 285, 89

\bibitem[{{Dr{\k{a}}{\.z}kowska} \& {Dullemond}(2014)}]{Draz14}
{Dr{\k{a}}{\.z}kowska}, J. \& {Dullemond}, C.~P. 2014, \aap, 572, A78

\bibitem[{{Dullemond} \& {Dominik}(2005)}]{Dullemond05}
{Dullemond}, C.~P. \& {Dominik}, C. 2005, \aap, 434, 971

\bibitem[{{Fitzgerald} {et~al.}(2007){Fitzgerald}, {Kalas}, {Duch{\^e}ne},
  {Pinte}, \& {Graham}}]{Fitzgerald07}
{Fitzgerald}, M.~P., {Kalas}, P.~G., {Duch{\^e}ne}, G., {Pinte}, C., \&
  {Graham}, J.~R. 2007, \apj, 670, 536

\bibitem[{{Fritscher} \& {Teiser}(2021)}]{Fritscher21}
{Fritscher}, M. \& {Teiser}, J. 2021, \apj, 923, 134

\bibitem[{{Fukagawa} {et~al.}(2006){Fukagawa}, {Tamura}, {Itoh}, {Kudo},
  {Imaeda}, {Oasa}, {Hayashi}, \& {Hayashi}}]{Fukagawa06}
{Fukagawa}, M., {Tamura}, M., {Itoh}, Y., {et~al.} 2006, \apjl, 636, L153

\bibitem[{{Graham} {et~al.}(2007){Graham}, {Kalas}, \& {Matthews}}]{Graham07}
{Graham}, J.~R., {Kalas}, P.~G., \& {Matthews}, B.~C. 2007, \apj, 654, 595

\bibitem[{{Gundlach} \& {Blum}(2015)}]{Gundlach15}
{Gundlach}, B. \& {Blum}, J. 2015, \apj, 798, 34

\bibitem[{{Gundlach} {et~al.}(2018){Gundlach}, {Schmidt}, {Kreuzig},
  {Bischoff}, {Rezaei}, {Kothe}, {Blum}, {Grzesik}, \& {Stoll}}]{Gundlach18}
{Gundlach}, B., {Schmidt}, K.~P., {Kreuzig}, C., {et~al.} 2018, \mnras, 479,
  1273

\bibitem[{{Gustafson} \& {Kolokolova}(1999)}]{Gustafson99}
{Gustafson}, B. {\r{A}}.~S. \& {Kolokolova}, L. 1999, \jgr, 104, 31711

\bibitem[{{Halder} {et~al.}(2018){Halder}, {Deb Roy}, \& {Das}}]{Halder18}
{Halder}, P., {Deb Roy}, P., \& {Das}, H.~S. 2018, \icarus, 312, 45

\bibitem[{{Hasegawa} {et~al.}(2021){Hasegawa}, {Suzuki}, {Tanaka}, {Kobayashi},
  \& {Wada}}]{Hasegawa21}
{Hasegawa}, Y., {Suzuki}, T.~K., {Tanaka}, H., {Kobayashi}, H., \& {Wada}, K.
  2021, \apj, 915, 22

\bibitem[{{Henning} \& {Stognienko}(1996)}]{Henning96}
{Henning}, T. \& {Stognienko}, R. 1996, \aap, 311, 291

\bibitem[{{Hull} {et~al.}(2018){Hull}, {Yang}, {Li}, {Kataoka}, {Stephens},
  {Andrews}, {Bai}, {Cleeves}, {Hughes}, {Looney}, {P{\'e}rez}, \&
  {Wilner}}]{Hull18}
{Hull}, C. L.~H., {Yang}, H., {Li}, Z.-Y., {et~al.} 2018, \apj, 860, 82

\bibitem[{{Hunziker} {et~al.}(2021){Hunziker}, {Schmid}, {Ma}, {Menard},
  {Avenhaus}, {Boccaletti}, {Beuzit}, {Chauvin}, {Dohlen}, {Dominik}, {Engler},
  {Ginski}, {Gratton}, {Henning}, {Langlois}, {Milli}, {Mouillet}, {Tschudi},
  {van Holstein}, \& {Vigan}}]{Hunziker21}
{Hunziker}, S., {Schmid}, H.~M., {Ma}, J., {et~al.} 2021, \aap, 648, A110

\bibitem[{{Johansen} {et~al.}(2007){Johansen}, {Oishi}, {Mac Low}, {Klahr},
  {Henning}, \& {Youdin}}]{Johansen07Nat}
{Johansen}, A., {Oishi}, J.~S., {Mac Low}, M.-M., {et~al.} 2007, \nat, 448,
  1022

\bibitem[{{Johansen} {et~al.}(2009){Johansen}, {Youdin}, \& {Mac
  Low}}]{Johansen09}
{Johansen}, A., {Youdin}, A., \& {Mac Low}, M.-M. 2009, \apjl, 704, L75

\bibitem[{{Jones} {et~al.}(2013){Jones}, {Fanciullo}, {K{\"o}hler},
  {Verstraete}, {Guillet}, {Bocchio}, \& {Ysard}}]{Jones13}
{Jones}, A.~P., {Fanciullo}, L., {K{\"o}hler}, M., {et~al.} 2013, \aap, 558,
  A62

\bibitem[{{Kataoka} {et~al.}(2013){Kataoka}, {Tanaka}, {Okuzumi}, \&
  {Wada}}]{Kataoka13}
{Kataoka}, A., {Tanaka}, H., {Okuzumi}, S., \& {Wada}, K. 2013, \aap, 557, L4

\bibitem[{{Kataoka} {et~al.}(2016){Kataoka}, {Tsukagoshi}, {Momose}, {Nagai},
  {Muto}, {Dullemond}, {Pohl}, {Fukagawa}, {Shibai}, {Hanawa}, \&
  {Murakawa}}]{Kataoka16}
{Kataoka}, A., {Tsukagoshi}, T., {Momose}, M., {et~al.} 2016, \apjl, 831, L12

\bibitem[{{Keller} \& {Messenger}(2011)}]{Keller11}
{Keller}, L.~P. \& {Messenger}, S. 2011, \gca, 75, 5336

\bibitem[{{Kimura}(2001)}]{Kimura01}
{Kimura}, H. 2001, \jqsrt, 70, 581

\bibitem[{{Kimura} {et~al.}(2016){Kimura}, {Kolokolova}, {Li}, \&
  {Lebreton}}]{Kimura16}
{Kimura}, H., {Kolokolova}, L., {Li}, A., \& {Lebreton}, J. 2016, arXiv
  e-prints, arXiv:1603.03123

\bibitem[{{Kimura} {et~al.}(2003){Kimura}, {Kolokolova}, \& {Mann}}]{Kimura03}
{Kimura}, H., {Kolokolova}, L., \& {Mann}, I. 2003, \aap, 407, L5

\bibitem[{{Kimura} {et~al.}(2006){Kimura}, {Kolokolova}, \& {Mann}}]{Kimura06}
{Kimura}, H., {Kolokolova}, L., \& {Mann}, I. 2006, \aap, 449, 1243

\bibitem[{{Kimura} \& {Mann}(2004)}]{Kimura04}
{Kimura}, H. \& {Mann}, I. 2004, \jqsrt, 89, 155

\bibitem[{{Kimura} {et~al.}(2020{\natexlab{a}}){Kimura}, {Wada}, {Kobayashi},
  {Senshu}, {Hirai}, {Yoshida}, {Kobayashi}, {Hong}, {Arai}, {Ishibashi}, \&
  {Yamada}}]{K20ice}
{Kimura}, H., {Wada}, K., {Kobayashi}, H., {et~al.} 2020{\natexlab{a}}, \mnras,
  498, 1801

\bibitem[{{Kimura} {et~al.}(2020{\natexlab{b}}){Kimura}, {Wada}, {Yoshida},
  {Hong}, {Senshu}, {Arai}, {Hirai}, {Kobayashi}, {Ishibashi}, \&
  {Yamada}}]{K20tensile}
{Kimura}, H., {Wada}, K., {Yoshida}, F., {et~al.} 2020{\natexlab{b}}, \mnras,
  496, 1667

\bibitem[{{Kobayashi} \& {Tanaka}(2021)}]{Kobayashi21}
{Kobayashi}, H. \& {Tanaka}, H. 2021, \apj, 922, 16

\bibitem[{{Kolokolova} \& {Kimura}(2010)}]{Kolokolova10}
{Kolokolova}, L. \& {Kimura}, H. 2010, \aap, 513, A40

\bibitem[{{Kolokolova} {et~al.}(2006){Kolokolova}, {Kimura}, {Ziegler}, \&
  {Mann}}]{Kolokolova06}
{Kolokolova}, L., {Kimura}, H., {Ziegler}, K., \& {Mann}, I. 2006, \jqsrt, 100,
  199

\bibitem[{{Kozasa} {et~al.}(1992){Kozasa}, {Blum}, \& {Mukai}}]{Kozasa92}
{Kozasa}, T., {Blum}, J., \& {Mukai}, T. 1992, \aap, 263, 423

\bibitem[{{Kozasa} {et~al.}(1993){Kozasa}, {Blum}, {Okamoto}, \&
  {Mukai}}]{Kozasa93}
{Kozasa}, T., {Blum}, J., {Okamoto}, H., \& {Mukai}, T. 1993, \aap, 276, 278

\bibitem[{{Krijt} {et~al.}(2016){Krijt}, {Ormel}, {Dominik}, \&
  {Tielens}}]{Krijt16}
{Krijt}, S., {Ormel}, C.~W., {Dominik}, C., \& {Tielens}, A.~G.~G.~M. 2016,
  \aap, 586, A20

\bibitem[{{Li} \& {Youdin}(2021)}]{Li21}
{Li}, R. \& {Youdin}, A.~N. 2021, \apj, 919, 107

\bibitem[{{Lorek} {et~al.}(2018){Lorek}, {Lacerda}, \& {Blum}}]{Lorek18}
{Lorek}, S., {Lacerda}, P., \& {Blum}, J. 2018, \aap, 611, A18

\bibitem[{{Lumme} \& {Penttil{\"a}}(2011)}]{Lumme11}
{Lumme}, K. \& {Penttil{\"a}}, A. 2011, \jqsrt, 112, 1658

\bibitem[{{Lumme} {et~al.}(1997){Lumme}, {Rahola}, \& {Hovenier}}]{Lumme97}
{Lumme}, K., {Rahola}, J., \& {Hovenier}, J.~W. 1997, \icarus, 126, 455

\bibitem[{{Ma} \& {Schmid}(2022)}]{Ma22}
{Ma}, J. \& {Schmid}, H.~M. 2022, arXiv e-prints, arXiv:2204.06370

\bibitem[{{Mackowski} \& {Mishchenko}(1996)}]{Mackowski96}
{Mackowski}, D.~W. \& {Mishchenko}, M.~I. 1996, Journal of the Optical Society
  of America A, 13, 2266

\bibitem[{{Mackowski} \& {Mishchenko}(2011)}]{Mackowski11}
{Mackowski}, D.~W. \& {Mishchenko}, M.~I. 2011, \jqsrt, 112, 2182

\bibitem[{{Mannel} {et~al.}(2019){Mannel}, {Bentley}, {Boakes}, {Jeszenszky},
  {Ehrenfreund}, {Engrand}, {Koeberl}, {Levasseur-Regourd}, {Romstedt},
  {Schmied}, {Torkar}, \& {Weber}}]{Mannel19}
{Mannel}, T., {Bentley}, M.~S., {Boakes}, P.~D., {et~al.} 2019, \aap, 630, A26

\bibitem[{{Mannel} {et~al.}(2016){Mannel}, {Bentley}, {Schmied}, {Jeszenszky},
  {Levasseur-Regourd}, {Romstedt}, \& {Torkar}}]{Mannel16}
{Mannel}, T., {Bentley}, M.~S., {Schmied}, R., {et~al.} 2016, \mnras, 462, S304

\bibitem[{{Markkanen} {et~al.}(2018){Markkanen}, {Agarwal}, {V{\"a}is{\"a}nen},
  {Penttil{\"a}}, \& {Muinonen}}]{Markkanen18}
{Markkanen}, J., {Agarwal}, J., {V{\"a}is{\"a}nen}, T., {Penttil{\"a}}, A., \&
  {Muinonen}, K. 2018, \apjl, 868, L16

\bibitem[{{Mathis} {et~al.}(1977){Mathis}, {Rumpl}, \& {Nordsieck}}]{MRN77}
{Mathis}, J.~S., {Rumpl}, W., \& {Nordsieck}, K.~H. 1977, \apj, 217, 425

\bibitem[{{Milli} {et~al.}(2017){Milli}, {Vigan}, {Mouillet}, {Lagrange},
  {Augereau}, {Pinte}, {Mawet}, {Schmid}, {Boccaletti}, {Matr{\`a}}, {Kral},
  {Ertel}, {Chauvin}, {Bazzon}, {M{\'e}nard}, {Beuzit}, {Thalmann}, {Dominik},
  {Feldt}, {Henning}, {Min}, {Girard}, {Galicher}, {Bonnefoy}, {Fusco}, {de
  Boer}, {Janson}, {Maire}, {Mesa}, {Schlieder}, \& {Sphere
  Consortium}}]{Milli17}
{Milli}, J., {Vigan}, A., {Mouillet}, D., {et~al.} 2017, \aap, 599, A108

\bibitem[{{Min} {et~al.}(2016){Min}, {Rab}, {Woitke}, {Dominik}, \&
  {M{\'e}nard}}]{Min16}
{Min}, M., {Rab}, C., {Woitke}, P., {Dominik}, C., \& {M{\'e}nard}, F. 2016,
  \aap, 585, A13

\bibitem[{{Monnier} {et~al.}(2019){Monnier}, {Harries}, {Bae}, {Setterholm},
  {Laws}, {Aarnio}, {Adams}, {Andrews}, {Calvet}, {Espaillat}, {Hartmann},
  {Kraus}, {McClure}, {Miller}, {Oppenheimer}, {Wilner}, \& {Zhu}}]{Monnier19}
{Monnier}, J.~D., {Harries}, T.~J., {Bae}, J., {et~al.} 2019, \apj, 872, 122

\bibitem[{{Moreno} {et~al.}(2007){Moreno}, {Mu{\~n}oz}, {Guirado}, \&
  {Vilaplana}}]{Moreno07}
{Moreno}, F., {Mu{\~n}oz}, O., {Guirado}, D., \& {Vilaplana}, R. 2007, \jqsrt,
  106, 348

\bibitem[{{Mori} \& {Kataoka}(2021)}]{Mori21}
{Mori}, T. \& {Kataoka}, A. 2021, \apj, 908, 153

\bibitem[{{Mu{\~n}oz} {et~al.}(2007){Mu{\~n}oz}, {Volten}, {Hovenier},
  {Nousiainen}, {Muinonen}, {Guirado}, {Moreno}, \& {Waters}}]{Munoz07}
{Mu{\~n}oz}, O., {Volten}, H., {Hovenier}, J.~W., {et~al.} 2007, Journal of
  Geophysical Research (Atmospheres), 112, D13215

\bibitem[{{Muinonen} {et~al.}(1996){Muinonen}, {Nousiainen}, {Fast}, {Lumme},
  \& {Peltoneimi}}]{Muinonen96}
{Muinonen}, K., {Nousiainen}, T., {Fast}, P., {Lumme}, K., \& {Peltoneimi}, J.
  1996, \jqsrt, 55, 577

\bibitem[{{Muinonen} \& {Pieniluoma}(2011)}]{Muinonen11}
{Muinonen}, K. \& {Pieniluoma}, T. 2011, \jqsrt, 112, 1747

\bibitem[{{Mukai} {et~al.}(1992){Mukai}, {Ishimoto}, {Kozasa}, {Blum}, \&
  {Greenberg}}]{Mukai92}
{Mukai}, T., {Ishimoto}, H., {Kozasa}, T., {Blum}, J., \& {Greenberg}, J.~M.
  1992, \aap, 262, 315

\bibitem[{{Murakawa} {et~al.}(2008){Murakawa}, {Oya}, {Pyo}, \&
  {Ishii}}]{Murakawa08}
{Murakawa}, K., {Oya}, S., {Pyo}, T.~S., \& {Ishii}, M. 2008, \aap, 492, 731

\bibitem[{{Musiolik} {et~al.}(2016{\natexlab{a}}){Musiolik}, {Teiser},
  {Jankowski}, \& {Wurm}}]{Musiolik16a}
{Musiolik}, G., {Teiser}, J., {Jankowski}, T., \& {Wurm}, G.
  2016{\natexlab{a}}, \apj, 818, 16

\bibitem[{{Musiolik} {et~al.}(2016{\natexlab{b}}){Musiolik}, {Teiser},
  {Jankowski}, \& {Wurm}}]{Musiolik16b}
{Musiolik}, G., {Teiser}, J., {Jankowski}, T., \& {Wurm}, G.
  2016{\natexlab{b}}, \apj, 827, 63

\bibitem[{{Musiolik} \& {Wurm}(2019)}]{Musiolik19}
{Musiolik}, G. \& {Wurm}, G. 2019, \apj, 873, 58

\bibitem[{{Nousiainen} {et~al.}(2003){Nousiainen}, {Muinonen}, \&
  {R{\"a}Is{\"a}Nen}}]{Nousiainen03}
{Nousiainen}, T., {Muinonen}, K., \& {R{\"a}Is{\"a}Nen}, P. 2003, Journal of
  Geophysical Research (Atmospheres), 108, 4025

\bibitem[{{Ohashi} \& {Kataoka}(2019)}]{Ohashi19}
{Ohashi}, S. \& {Kataoka}, A. 2019, \apj, 886, 103

\bibitem[{{Ohashi} {et~al.}(2018){Ohashi}, {Kataoka}, {Nagai}, {Momose},
  {Muto}, {Hanawa}, {Fukagawa}, {Tsukagoshi}, {Murakawa}, \&
  {Shibai}}]{Ohashi18}
{Ohashi}, S., {Kataoka}, A., {Nagai}, H., {et~al.} 2018, \apj, 864, 81

\bibitem[{{Ohashi} {et~al.}(2020){Ohashi}, {Kataoka}, {van der Marel}, {Hull},
  {Dent}, {Pohl}, {Pinilla}, {van Dishoeck}, \& {Henning}}]{Ohashi20}
{Ohashi}, S., {Kataoka}, A., {van der Marel}, N., {et~al.} 2020, \apj, 900, 81

\bibitem[{{Okuzumi} {et~al.}(2016){Okuzumi}, {Momose}, {Sirono}, {Kobayashi},
  \& {Tanaka}}]{Okuzumi16}
{Okuzumi}, S., {Momose}, M., {Sirono}, S.-i., {Kobayashi}, H., \& {Tanaka}, H.
  2016, \apj, 821, 82

\bibitem[{{Okuzumi} {et~al.}(2012){Okuzumi}, {Tanaka}, {Kobayashi}, \&
  {Wada}}]{Okuzumi12}
{Okuzumi}, S., {Tanaka}, H., {Kobayashi}, H., \& {Wada}, K. 2012, \apj, 752,
  106

\bibitem[{{Okuzumi} \& {Tazaki}(2019)}]{Okuzumi19}
{Okuzumi}, S. \& {Tazaki}, R. 2019, \apj, 878, 132

\bibitem[{{Ormel} {et~al.}(2007){Ormel}, {Spaans}, \& {Tielens}}]{Ormel07}
{Ormel}, C.~W., {Spaans}, M., \& {Tielens}, A.~G.~G.~M. 2007, \aap, 461, 215

\bibitem[{{Perrin} {et~al.}(2015){Perrin}, {Duchene}, {Millar-Blanchaer},
  {Fitzgerald}, {Graham}, {Wiktorowicz}, {Kalas}, {Macintosh}, {Bauman},
  {Cardwell}, {Chilcote}, {De Rosa}, {Dillon}, {Doyon}, {Dunn}, {Erikson},
  {Gavel}, {Goodsell}, {Hartung}, {Hibon}, {Ingraham}, {Kerley}, {Konapacky},
  {Larkin}, {Maire}, {Marchis}, {Marois}, {Mittal}, {Morzinski}, {Oppenheimer},
  {Palmer}, {Patience}, {Poyneer}, {Pueyo}, {Rantakyr{\"o}}, {Sadakuni},
  {Saddlemyer}, {Savransky}, {Soummer}, {Sivaramakrishnan}, {Song}, {Thomas},
  {Wallace}, {Wang}, \& {Wolff}}]{Perrin15}
{Perrin}, M.~D., {Duchene}, G., {Millar-Blanchaer}, M., {et~al.} 2015, \apj,
  799, 182

\bibitem[{{Perrin} {et~al.}(2009){Perrin}, {Schneider}, {Duchene}, {Pinte},
  {Grady}, {Wisniewski}, \& {Hines}}]{Perrin09}
{Perrin}, M.~D., {Schneider}, G., {Duchene}, G., {et~al.} 2009, \apjl, 707,
  L132

\bibitem[{{Petrova} {et~al.}(2000){Petrova}, {Jockers}, \&
  {Kiselev}}]{Petrova00}
{Petrova}, E.~V., {Jockers}, K., \& {Kiselev}, N.~N. 2000, \icarus, 148, 526

\bibitem[{{Petrova} {et~al.}(2004){Petrova}, {Tishkovets}, \&
  {Jockers}}]{Petrova04}
{Petrova}, E.~V., {Tishkovets}, V.~P., \& {Jockers}, K. 2004, Solar System
  Research, 38, 309

\bibitem[{{Pinilla} {et~al.}(2017){Pinilla}, {Pohl}, {Stammler}, \&
  {Birnstiel}}]{Pinilla17}
{Pinilla}, P., {Pohl}, A., {Stammler}, S.~M., \& {Birnstiel}, T. 2017, \apj,
  845, 68

\bibitem[{{Pollack} {et~al.}(1994){Pollack}, {Hollenbach}, {Beckwith},
  {Simonelli}, {Roush}, \& {Fong}}]{Pollack94}
{Pollack}, J.~B., {Hollenbach}, D., {Beckwith}, S., {et~al.} 1994, \apj, 421,
  615

\bibitem[{{Poppe} {et~al.}(2000){Poppe}, {Blum}, \& {Henning}}]{Poppe00}
{Poppe}, T., {Blum}, J., \& {Henning}, T. 2000, \apj, 533, 454

\bibitem[{{Poteet} {et~al.}(2018){Poteet}, {Chen}, {Hines}, {Perrin}, {Debes},
  {Pueyo}, {Schneider}, {Mazoyer}, \& {Kolokolova}}]{Poteet18}
{Poteet}, C.~A., {Chen}, C.~H., {Hines}, D.~C., {et~al.} 2018, \apj, 860, 115

\bibitem[{{Rameau} {et~al.}(2012){Rameau}, {Chauvin}, {Lagrange},
  {Th{\'e}bault}, {Milli}, {Girard}, \& {Bonnefoy}}]{Rameau12}
{Rameau}, J., {Chauvin}, G., {Lagrange}, A.~M., {et~al.} 2012, \aap, 546, A24

\bibitem[{{Rietmeijer}(1993)}]{Rietmeijer93}
{Rietmeijer}, F.~J.~M. 1993, Earth and Planetary Science Letters, 117, 609

\bibitem[{{Rodigas} {et~al.}(2014){Rodigas}, {Follette}, {Weinberger}, {Close},
  \& {Hines}}]{Rodigas14}
{Rodigas}, T.~J., {Follette}, K.~B., {Weinberger}, A., {Close}, L., \& {Hines},
  D.~C. 2014, \apjl, 791, L37

\bibitem[{{Seizinger} \& {Kley}(2013)}]{Seizinger13}
{Seizinger}, A. \& {Kley}, W. 2013, \aap, 551, A65

\bibitem[{{Sekiya} \& {Onishi}(2018)}]{Sekiya18}
{Sekiya}, M. \& {Onishi}, I.~K. 2018, \apj, 860, 140

\bibitem[{{Shen} {et~al.}(2008){Shen}, {Draine}, \& {Johnson}}]{Shen08}
{Shen}, Y., {Draine}, B.~T., \& {Johnson}, E.~T. 2008, \apj, 689, 260

\bibitem[{{Shen} {et~al.}(2009){Shen}, {Draine}, \& {Johnson}}]{Shen09}
{Shen}, Y., {Draine}, B.~T., \& {Johnson}, E.~T. 2009, \apj, 696, 2126

\bibitem[{{Silber} {et~al.}(2000){Silber}, {Gledhill}, {Duch{\^e}ne}, \&
  {M{\'e}nard}}]{Silber00}
{Silber}, J., {Gledhill}, T., {Duch{\^e}ne}, G., \& {M{\'e}nard}, F. 2000,
  \apjl, 536, L89

\bibitem[{{Stephens} {et~al.}(2017){Stephens}, {Yang}, {Li}, {Looney},
  {Kataoka}, {Kwon}, {Fern{\'a}ndez-L{\'o}pez}, {Hull}, {Hughes}, {Segura-Cox},
  {Mundy}, {Crutcher}, \& {Rao}}]{Stephens17}
{Stephens}, I.~W., {Yang}, H., {Li}, Z.-Y., {et~al.} 2017, \apj, 851, 55

\bibitem[{{Tanii} {et~al.}(2012){Tanii}, {Itoh}, {Kudo}, {Hioki}, {Oasa},
  {Gupta}, {Sen}, {Wisniewski}, {Muto}, {Grady}, {Hashimoto}, {Fukagawa},
  {Mayama}, {Hornbeck}, {Sitko}, {Russell}, {Werren}, {Cur{\'e}}, {Currie},
  {Ohashi}, {Okamoto}, {Momose}, {Honda}, {Inutsuka}, {Takeuchi}, {Dong},
  {Abe}, {Brandner}, {Brandt}, {Carson}, {Egner}, {Feldt}, {Fukue}, {Goto},
  {Guyon}, {Hayano}, {Hayashi}, {Hayashi}, {Henning}, {Hodapp}, {Ishii}, {Iye},
  {Janson}, {Kandori}, {Knapp}, {Kusakabe}, {Kuzuhara}, {Matsuo}, {McElwain},
  {Miyama}, {Morino}, {Moro-Mart{\'\i}n}, {Nishimura}, {Pyo}, {Serabyn},
  {Suto}, {Suzuki}, {Takami}, {Takato}, {Terada}, {Thalmann}, {Tomono},
  {Turner}, {Watanabe}, {Yamada}, {Takami}, {Usuda}, \& {Tamura}}]{Tanii12}
{Tanii}, R., {Itoh}, Y., {Kudo}, T., {et~al.} 2012, \pasj, 64, 124

\bibitem[{{Tatsuuma} {et~al.}(2019){Tatsuuma}, {Kataoka}, \&
  {Tanaka}}]{Tatsuuma19}
{Tatsuuma}, M., {Kataoka}, A., \& {Tanaka}, H. 2019, \apj, 874, 159

\bibitem[{{Tazaki} {et~al.}(2021){Tazaki}, {Murakawa}, {Muto}, {Honda}, \&
  {Inoue}}]{Tazaki21}
{Tazaki}, R., {Murakawa}, K., {Muto}, T., {Honda}, M., \& {Inoue}, A.~K. 2021,
  \apj, 921, 173

\bibitem[{{Tazaki} \& {Tanaka}(2018)}]{Tazaki18}
{Tazaki}, R. \& {Tanaka}, H. 2018, \apj, 860, 79

\bibitem[{{Tazaki} {et~al.}(2019){Tazaki}, {Tanaka}, {Muto}, {Kataoka}, \&
  {Okuzumi}}]{Tazaki19}
{Tazaki}, R., {Tanaka}, H., {Muto}, T., {Kataoka}, A., \& {Okuzumi}, S. 2019,
  \mnras, 485, 4951

\bibitem[{{Tazaki} {et~al.}(2016){Tazaki}, {Tanaka}, {Okuzumi}, {Kataoka}, \&
  {Nomura}}]{Tazaki16}
{Tazaki}, R., {Tanaka}, H., {Okuzumi}, S., {Kataoka}, A., \& {Nomura}, H. 2016,
  \apj, 823, 70

\bibitem[{{Tschudi} \& {Schmid}(2021)}]{Tschudi21}
{Tschudi}, C. \& {Schmid}, H.~M. 2021, \aap, 655, A37

\bibitem[{{Volten} {et~al.}(2007){Volten}, {Mu{\~n}oz}, {Hovenier},
  {Rietmeijer}, {Nuth}, {Waters}, \& {van der Zande}}]{Volten07}
{Volten}, H., {Mu{\~n}oz}, O., {Hovenier}, J.~W., {et~al.} 2007, \aap, 470, 377

\bibitem[{{Wada} {et~al.}(2013){Wada}, {Tanaka}, {Okuzumi}, {Kobayashi},
  {Suyama}, {Kimura}, \& {Yamamoto}}]{Wada13}
{Wada}, K., {Tanaka}, H., {Okuzumi}, S., {et~al.} 2013, \aap, 559, A62

\bibitem[{{Wada} {et~al.}(2009){Wada}, {Tanaka}, {Suyama}, {Kimura}, \&
  {Yamamoto}}]{Wada09}
{Wada}, K., {Tanaka}, H., {Suyama}, T., {Kimura}, H., \& {Yamamoto}, T. 2009,
  \apj, 702, 1490

\bibitem[{{Wada} {et~al.}(2011){Wada}, {Tanaka}, {Suyama}, {Kimura}, \&
  {Yamamoto}}]{Wada11}
{Wada}, K., {Tanaka}, H., {Suyama}, T., {Kimura}, H., \& {Yamamoto}, T. 2011,
  \apj, 737, 36

\bibitem[{{Warren} \& {Brandt}(2008)}]{Warren08}
{Warren}, S.~G. \& {Brandt}, R.~E. 2008, Journal of Geophysical Research
  (Atmospheres), 113, D14220

\bibitem[{{Weidenschilling}(1984)}]{Weidenschilling84}
{Weidenschilling}, S.~J. 1984, \icarus, 60, 553

\bibitem[{{Weidenschilling}(1997)}]{Weidenschilling97}
{Weidenschilling}, S.~J. 1997, \icarus, 127, 290

\bibitem[{{Weidling} {et~al.}(2009){Weidling}, {G{\"u}ttler}, {Blum}, \&
  {Brauer}}]{Weidling09}
{Weidling}, R., {G{\"u}ttler}, C., {Blum}, J., \& {Brauer}, F. 2009, \apj, 696,
  2036

\bibitem[{{West}(1991)}]{West91}
{West}, R.~A. 1991, \ao, 30, 5316

\bibitem[{{Wozniakiewicz} {et~al.}(2013){Wozniakiewicz}, {Bradley}, {Ishii},
  {Price}, \& {Brownlee}}]{Woz13}
{Wozniakiewicz}, P.~J., {Bradley}, J.~P., {Ishii}, H.~A., {Price}, M.~C., \&
  {Brownlee}, D.~E. 2013, \apj, 779, 164

\bibitem[{{Xing} \& {Hanner}(1997)}]{Xing97}
{Xing}, Z. \& {Hanner}, M.~S. 1997, \aap, 324, 805

\bibitem[{{Yanamandra-Fisher} \& {Hanner}(1999)}]{Yanamandra99}
{Yanamandra-Fisher}, P.~A. \& {Hanner}, M.~S. 1999, \icarus, 138, 107

\bibitem[{{Yang} {et~al.}(2017){Yang}, {Johansen}, \& {Carrera}}]{Yang17}
{Yang}, C.-C., {Johansen}, A., \& {Carrera}, D. 2017, \aap, 606, A80

\bibitem[{{Youdin} \& {Goodman}(2005)}]{Youdin05}
{Youdin}, A.~N. \& {Goodman}, J. 2005, \apj, 620, 459

\bibitem[{{Yurkin} \& {Hoekstra}(2011)}]{Yurkin11}
{Yurkin}, M.~A. \& {Hoekstra}, A.~G. 2011, \jqsrt, 112, 2234

\bibitem[{{Zerull} {et~al.}(1993){Zerull}, {Gustafson}, {Schulz}, \&
  {Thiele-Corbach}}]{Zerull93}
{Zerull}, R.~H., {Gustafson}, B.~A.~S., {Schulz}, K., \& {Thiele-Corbach}, E.
  1993, \ao, 32, 4088

\bibitem[{{Zsom} {et~al.}(2010){Zsom}, {Ormel}, {G{\"u}ttler}, {Blum}, \&
  {Dullemond}}]{Zsom10}
{Zsom}, A., {Ormel}, C.~W., {G{\"u}ttler}, C., {Blum}, J., \& {Dullemond},
  C.~P. 2010, \aap, 513, A57

\bibitem[{{Zubko} {et~al.}(1996){Zubko}, {Mennella}, {Colangeli}, \&
  {Bussoletti}}]{Zubko96}
{Zubko}, V.~G., {Mennella}, V., {Colangeli}, L., \& {Bussoletti}, E. 1996,
  \mnras, 282, 1321

\end{thebibliography}

\begin{appendix}

\section{Maximum polarization of BAM1 and BAM2 clusters} \label{sec:appsize}

\begin{figure*}[t]
\begin{center}
\includegraphics[width=0.99\linewidth]{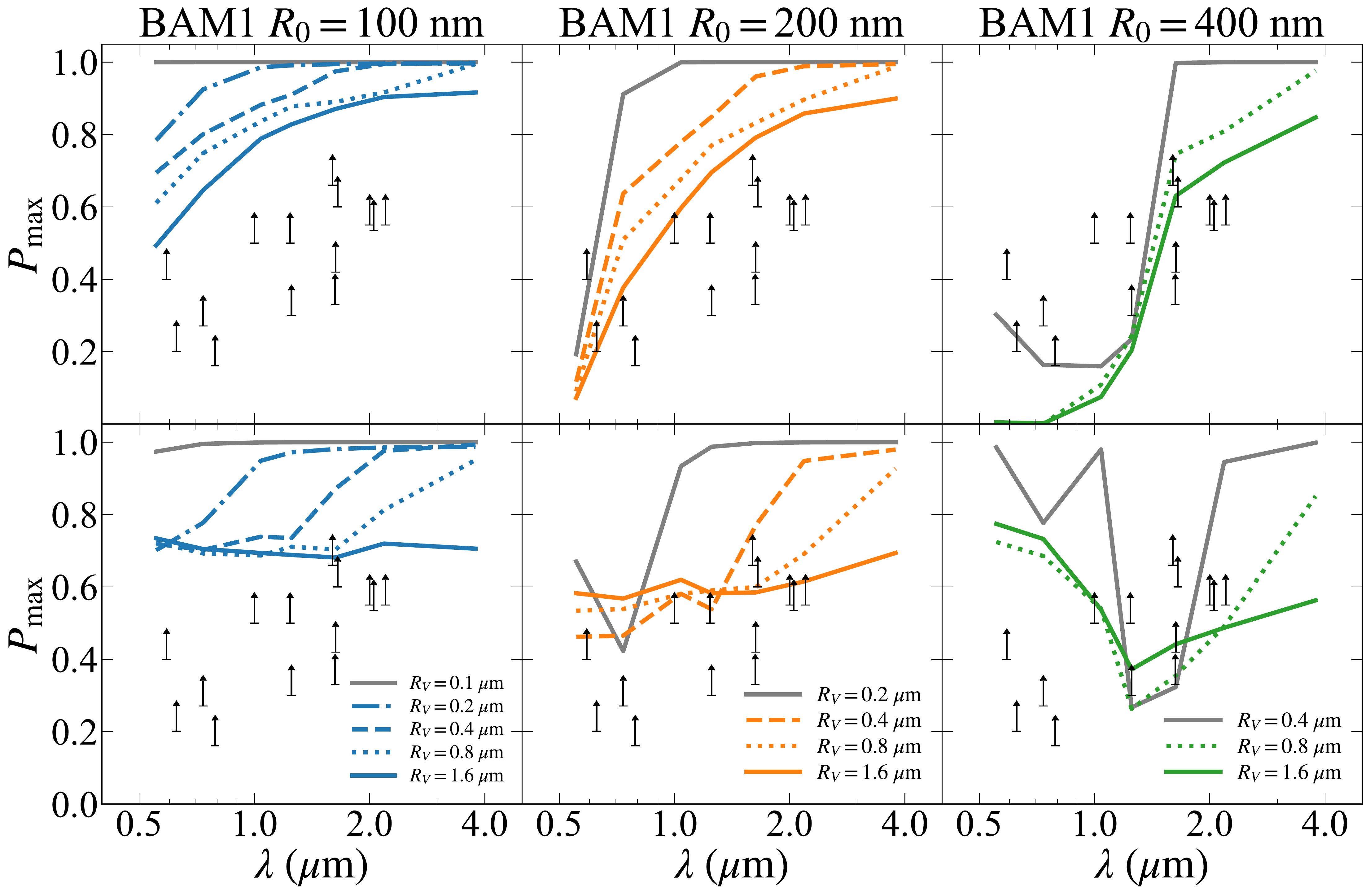}
\includegraphics[width=0.99\linewidth]{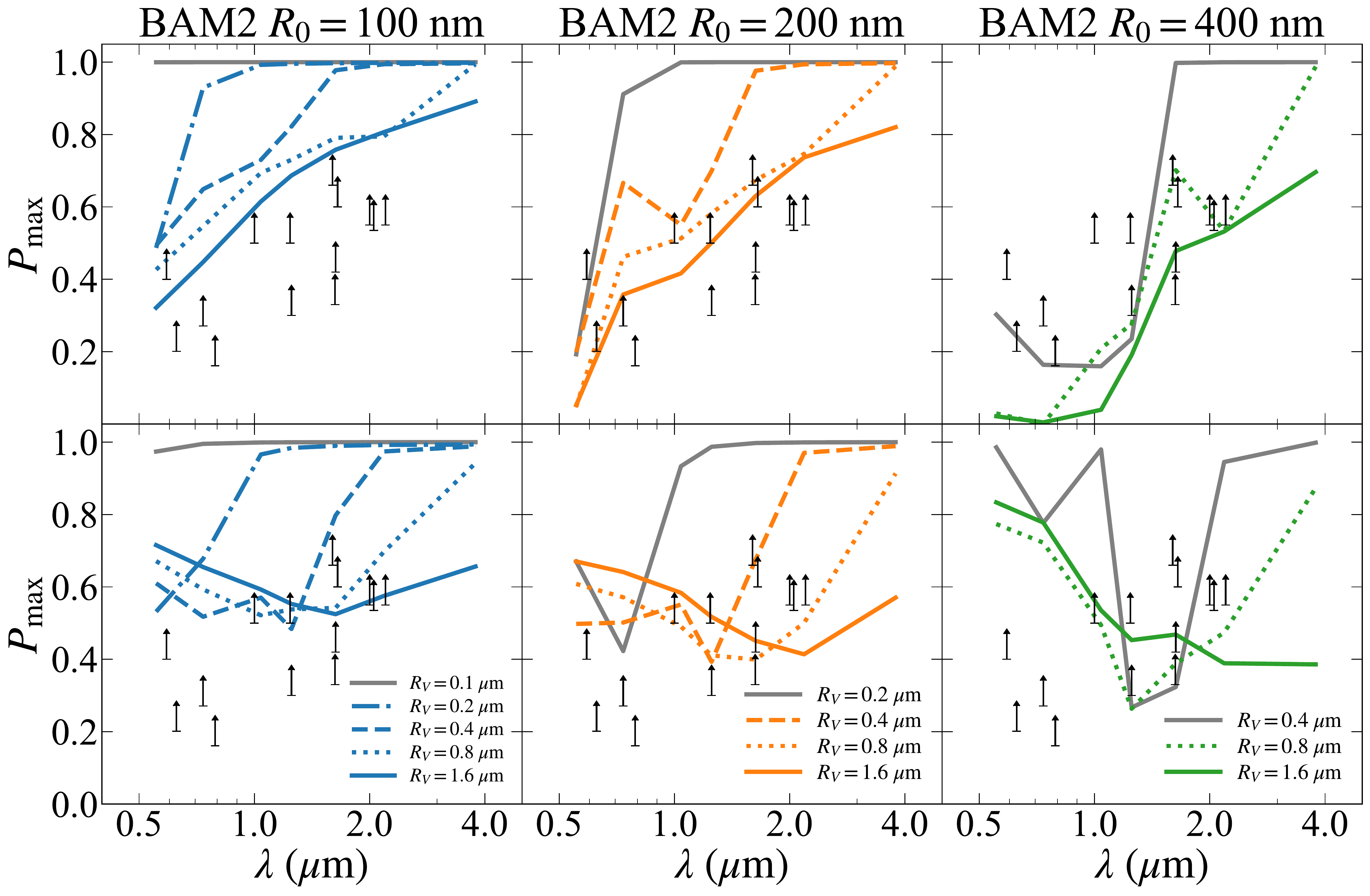}
\caption{Same as Figure \ref{fig:bpcaall}, but for BAM1 ({\it upper panel}) and for BAM2  ({\it lower panel}).}
\label{fig:bamall}
\end{center}
\end{figure*}

Fig. \ref{fig:bamall} shows the maximum polarization for BAM1 and BAM2 clusters.

\section{Effect of particle shapes on the maximum polarization} \label{sec:shape}

This study assumes spherical monomers. One of the artifacts arising from the assumption of spherical monomers is appearance of oscillatory patterns, sometimes referred to as ripples, in polarization curves when $\xmon>1$ \citep{Bohren83}. Irregular particle shape usually results in smoother polarization curves, particularly for absorbing materials \citep{Yanamandra99, Muinonen11}. Although the presence of touching spherical monomers in an aggregate tends to suppress the oscillation \citep{Xing97}, it is still possible that the perfect sphere assumption might introduce some artifact in wavelength dependence of polarization, particularly when the monomer size parameter exceeds unity.

Previous studies show that the maximum polarization of aggregates of nonspherical monomers tends to be similar or slightly lowered to those of spherical monomers. \citet{Xing97} compared optical properties of aggregates of spheres and tetrahedral monomers with $m=1.88+0.71i$. Although the scattering polarization by a single sphere and tetrahedron differs significantly, once the monomers are aggregated and touched each other, the resultant maximum polarizations from aggregates of 10-spheres and -tetrahedrons are surprisingly similar to each other \citep [see, e.g., Fig. 9ab in][]{Xing97}.
Meanwhile, \citet{Moreno07} investigated aggregates of 256-spheres and -cubes with an absorbing composition and found that $\pmax$ of aggregates of cubic monomers show a lower polarization degree by about 20\% points. In their comparison, the side length of the cube and sphere diameter were taken to be the same. However, as pointed out by \citet{Kimura16}, their comparison allows cubic monomers to be larger than spherical monomers in terms of volume, and thus it is unclear to what extent the difference in $\pmax$ can be purely attributed to the shape effect of the monomers. \citet{Lumme11} performed simulations using discrete dipole approximation (DDA) for aggregates consisting of Gaussian random sphere (GRS) particles with a less absorbing composition ($k=0.01$). Although the maximum polarization degree is lowered for aggregates of nonspherical monomers, the effect is found to be non-significant.

\begin{figure}[t]
\begin{center}
\includegraphics[width=0.99\linewidth]{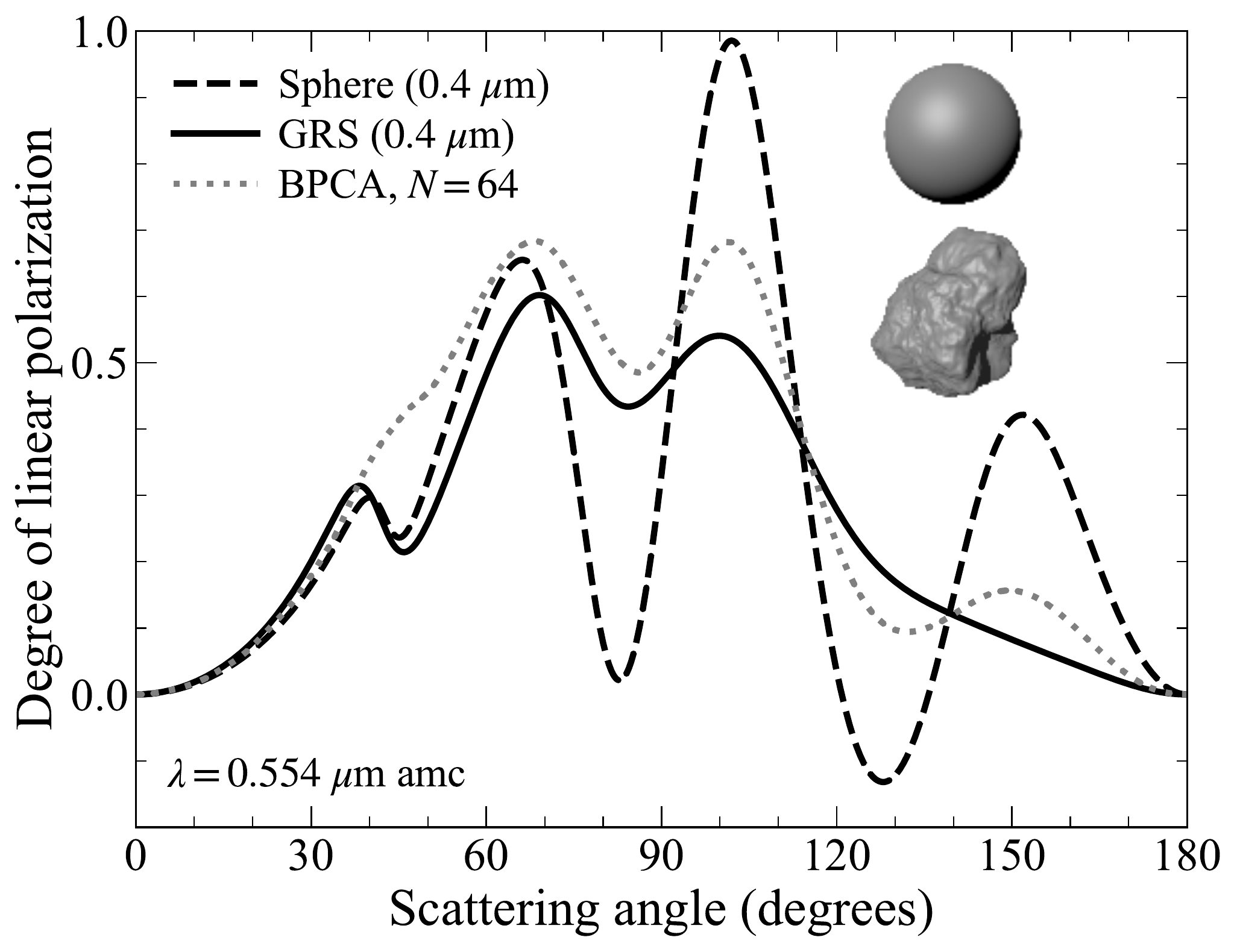}
\caption{Effect of non-sphericity on polarization curves. Scattered light of a single sphere (dashed line) and GRS particles (solid line) with \amc-\texttt{400} at $\lambda=0.554~\mu$m. The volume-equivalent size parameter is 4.54. The results for GRS particles are averaged over 10 realizations. A realization of GRS particles and a sphere are also shown.  For comparison, the polarization curve for aggregates with 64-spherical monomers with \amc-\texttt{400} is shown as well.}
\label{fig:pangshape}
\end{center}
\end{figure}

\begin{figure}[t]
\begin{center}
\includegraphics[width=0.99\linewidth]{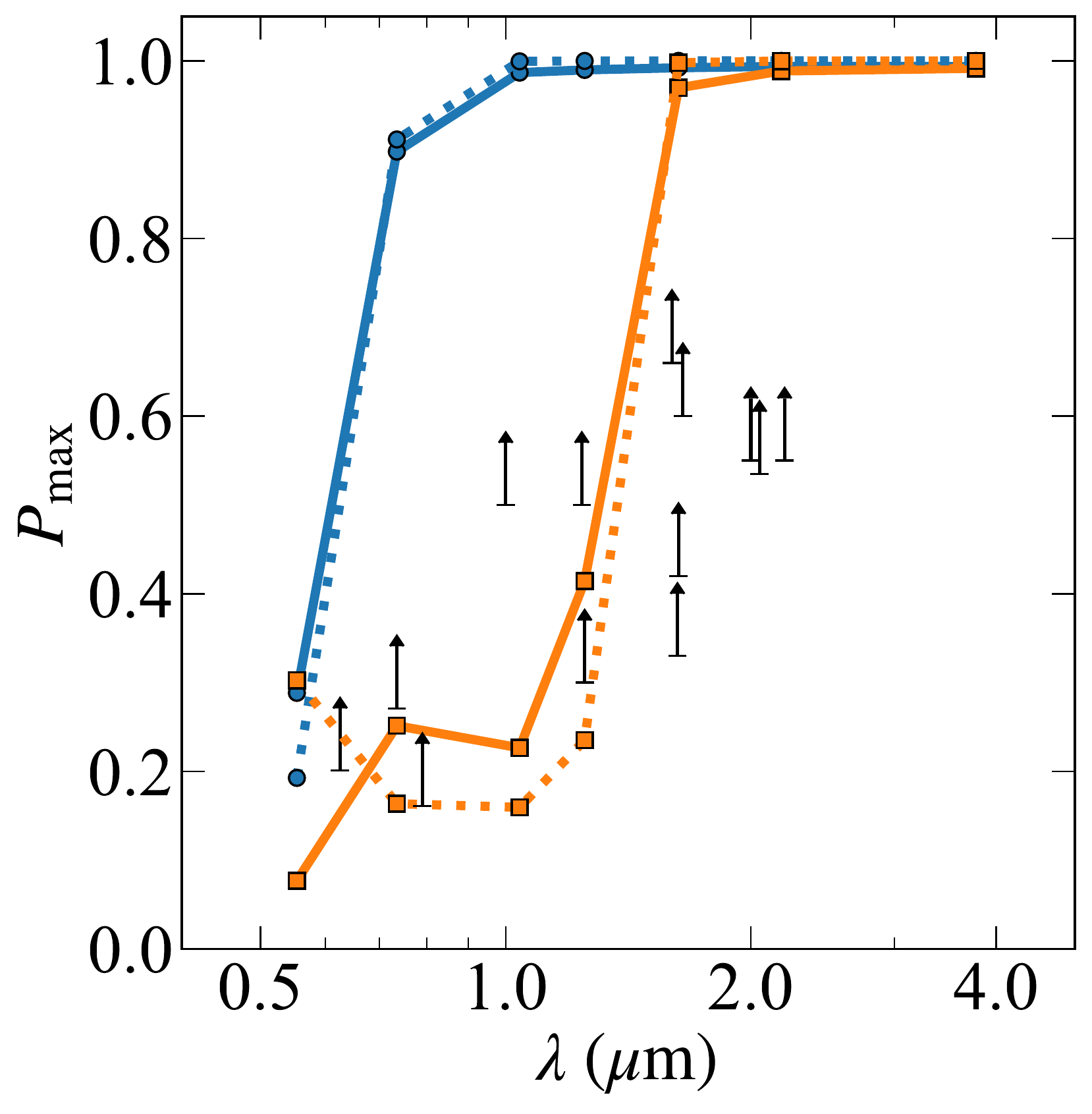}
\includegraphics[width=0.99\linewidth]{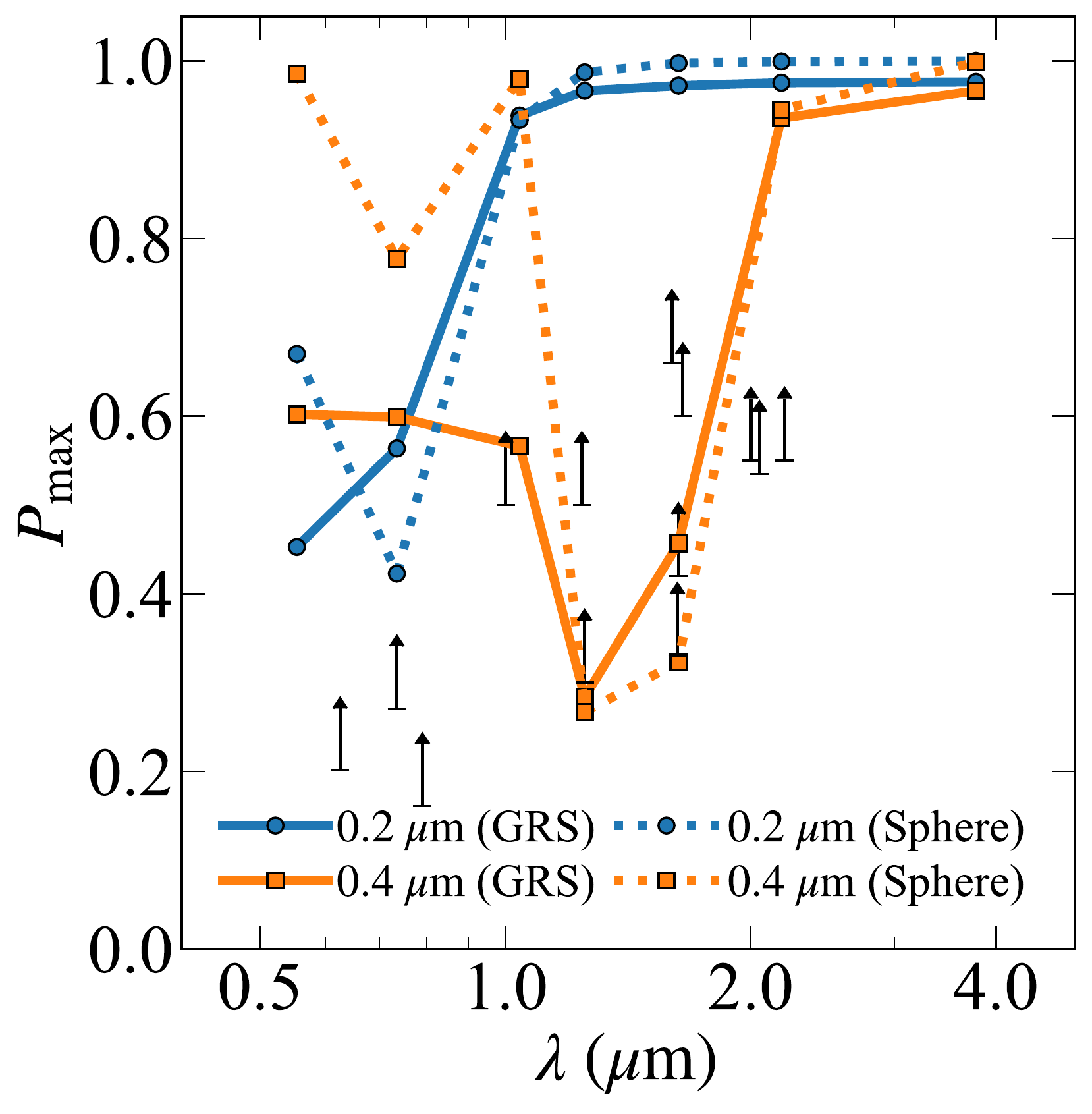}
\caption{Effect of particle shape on the degree of polarization for the \org~({\it upper panel}) and \amc~({\it lower panel}) compositions. The blue and orange lines represent the results for $\rv=0.2~\mu$m and $0.4~\mu$m, respectively. The solid and dashed lines represent the results for GRS particles and spherical particles, respectively.}
\label{fig:shape}
\end{center}
\end{figure}

To assess the effect of monomer shape on the degree of polarization, we compare scattering polarization of a single sphere and a nonspherical particle. As a nonspherical particle, we generated GRS particles \citep{Muinonen96}. As a Legendre expansion coefficient for the autocorrelation function, we adopt a power-law function $c_l\propto l^{-\nu}$ ($l\ge2$) \citep{Nousiainen03}. We adopt $\nu=3.4$ and the relative standard deviation of radius $\sigma=0.2$. With these parameters, we can nicely mimic irregularity of solid particles in nature, i.e., saharan dust particles \citep{Munoz07}. We generated 10 realizations of GRS particles. One realization of GRS particles is shown in Fig. \ref{fig:pangshape}. We adopt a DDA technique to solve the light scattering by GRS particles, using a publicly available code \texttt{ADDA}  \citep{Yurkin11}. Since the impact of non-sphericity on the degree of polarization appears mainly for $\xmon\gtrsim1$, we consider particles with $200$ nm and $400$ nm in volume-equivalent radius.

The results are shown in Fig. \ref{fig:shape}. At near-IR wavelengths, the maximum degree of polarization of the GRS particles is almost identical to that of spherical particles for all cases. On the other hand, at optical wavelengths, the maximum polarization of the GRS particles deviates significantly from that of the spherical particles for the \amc~composition. This is mainly because the nonspherical particles smear out ripple patterns in the polarization curves (Fig. \ref{fig:pangshape}). 

Because such a strong ripple can be similarly suppressed by the presence of touching spherical monomers (see the polarization curve of the BPCA clusters in Fig. \ref{fig:pangshape}), the differences in the maximum polarization between aggregates made of GRS and spherical particles would be less noticeable than those seen in Fig. \ref{fig:shape} \citep{Xing97}. However, some difference may possibly remain. Therefore, the assumption of spherical monomers would be robust at least around near-IR wavelengths, but might introduce some artifacts at optical wavelengths, particularly for \amc-\texttt{200} and -\texttt{400}. 

\section{Effect of an intermediate composition on the maximum polarization} \label{sec:intermed}
\begin{figure*}[t]
\begin{center}
\includegraphics[width=0.49\linewidth]{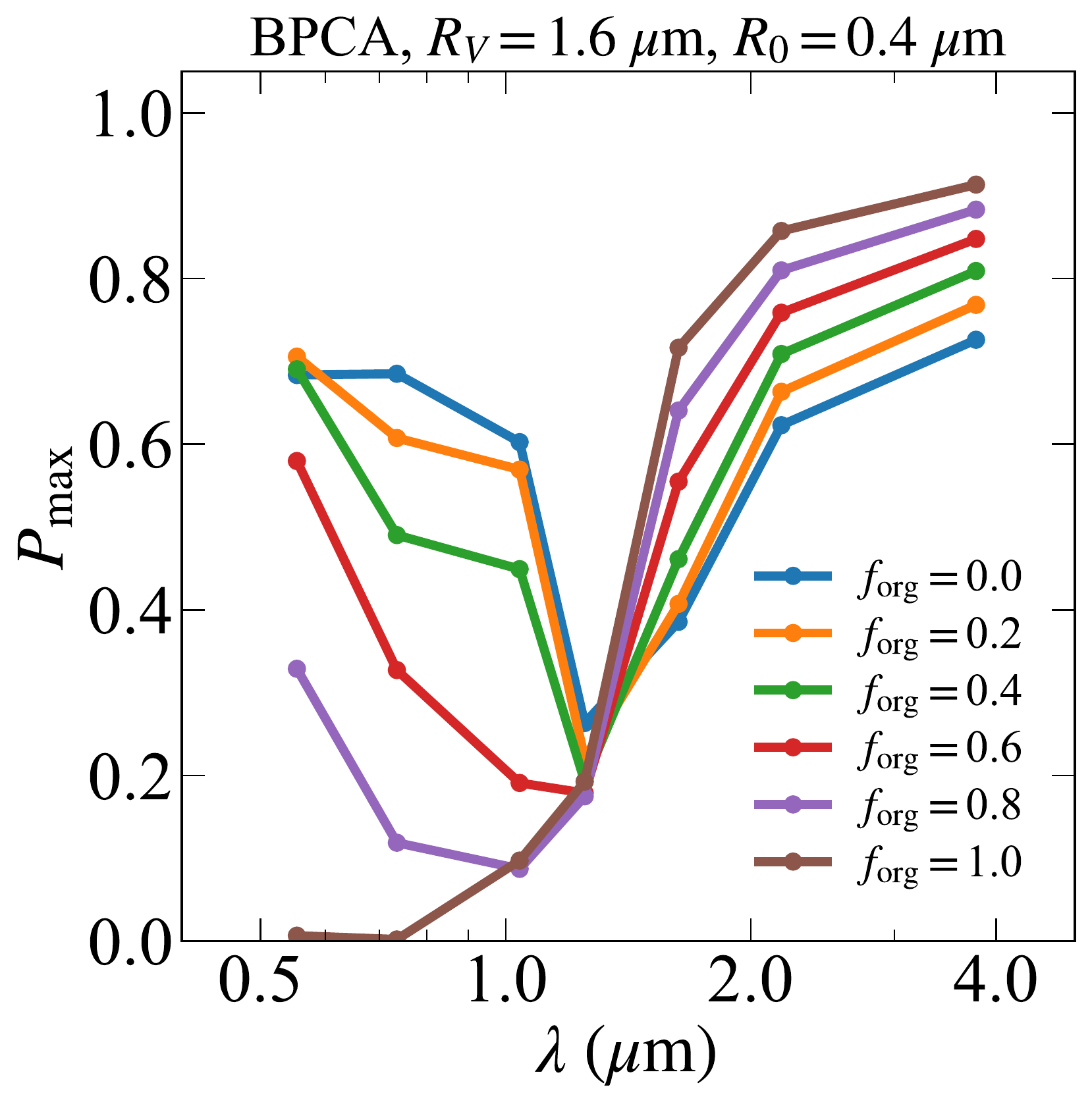}
\includegraphics[width=0.49\linewidth]{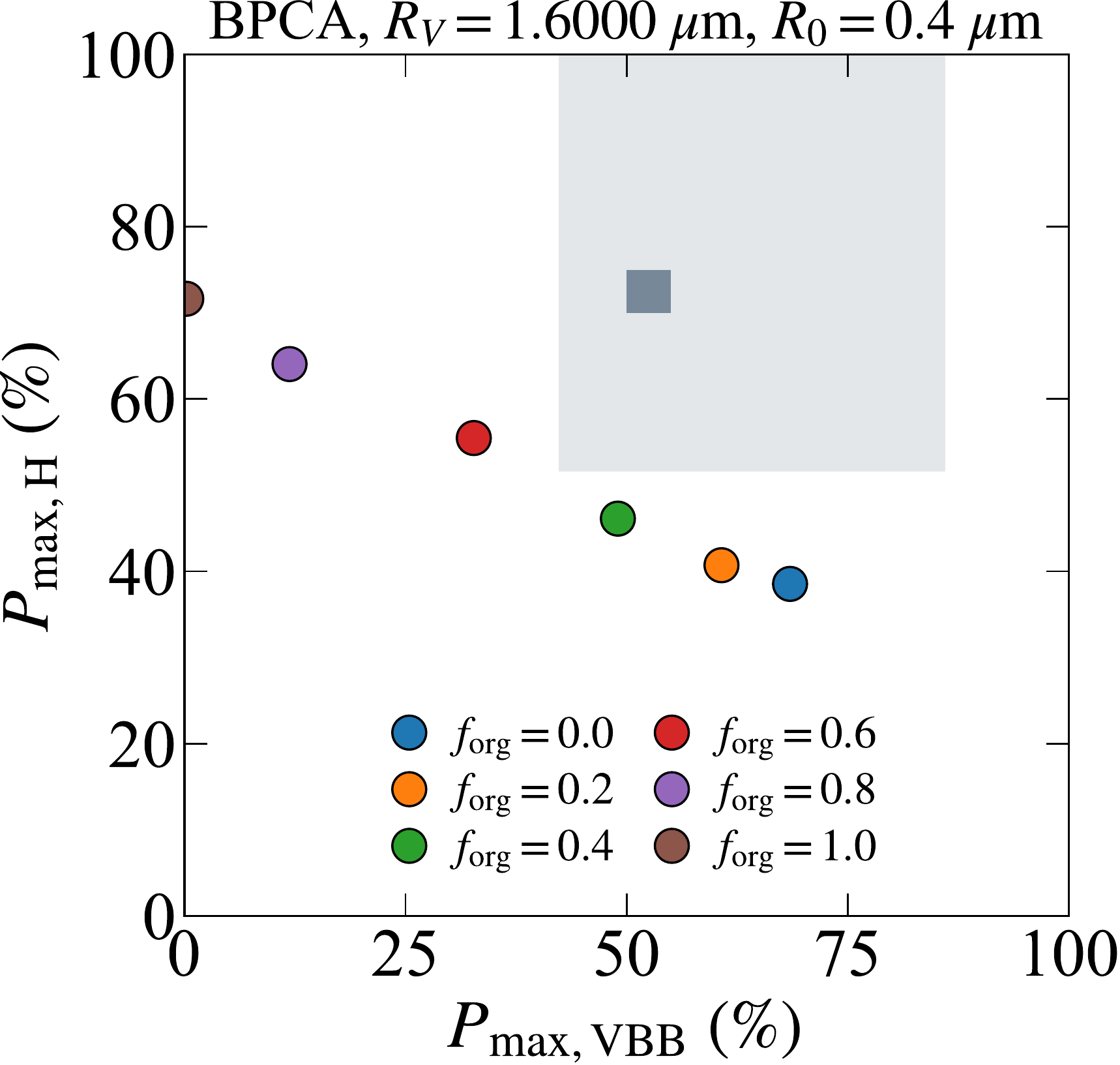}
\caption{{\it Left panel}: BPCA clusters with $\rv=1.6~\mu$m and $\rmon=400$ nm with 5-composite materials (silicate, water ice, amorphous carbon, organics, and troilite). $f_\mathrm{org}$ represents a mass fraction of organics with respect to total carbonaceous materials (organics + amorphous). $f_\mathrm{org}=0.0$ and $1.0$ mean the carbonaceous component is pure amorphous carbon and organics, respectively. {\it Right panel}: Same as Fig. \ref{fig:hd142527}, but for BPCA clusters with $\rv=1.6~\mu$m and $\rmon=400$ nm with 5-composite materials.}
\label{fig:opcont}
\end{center}
\end{figure*}

In Sect. \ref{sec:142527}, we argued that the observed maximum polarization fractions are inconsistent with aggregates with $\rmon=400$ nm. However, it is unclear to what extent this conclusion remains true if we relax the assumption of refractive indices. In particular, it follows from Fig. \ref{fig:hd142527} that an intermediate composition between \org~ and \amc~copmosition may yield a closer fit to the observations.

To test this possibility, we performed additional $T$-matrix simulations for BPCA clusters with an intermediate dust composition. We now consider a mixture of organics and amorphous carbon as a carbonaceous component so that each monomer is made of 5 different materials: silicate, water ice, organics, amorphous carbon, and troilite. By changing the mass ratio of organics to total carbon (organics + amorphous carbon), we calculated the maximum degree of polarization.

The results are shown in Fig. \ref{fig:opcont}. It is found that even if we consider an intermediate composition, the results are still located outside the observationally inferred region. Although models with $f_\mathrm{org}=0.4$--$0.6$ may be close to the observation, these models predict a significantly low polarization fraction at the $J$-band (Fig. \ref{fig:opcont}, left). The presence/absence of such a dip in the polarization fraction can be testable by future observations.

\end{appendix}

\end{document}